\documentclass[aps,
superscriptaddress,
showpacs,
preprintnumbers,
nofootinbib,
bibnotes,
amsmath,
amssymb,
twocolumn,
prl,
nosort,
floatfix,
]{revtex4-2}
\usepackage[utf8]{inputenc}
\usepackage{array}
\usepackage[normalem]{ulem}
\usepackage{amsfonts} 
\usepackage{amssymb} 
\usepackage{amsmath} 
\usepackage{graphicx} 
\usepackage{subfigure} 
\usepackage{array} 
\usepackage{dcolumn} 
\usepackage{latexsym} 
\usepackage{longtable} 
\usepackage{bbold}
\usepackage{color}
\usepackage{xcolor}
\usepackage{multirow}
\usepackage{rotating}
\usepackage{comment}
\usepackage{textgreek}
\usepackage[unicode]{hyperref} 

\newcommand{\bx}{{\textbf{x}}}

\usepackage[unicode]{hyperref} 
\hypersetup{
  colorlinks = true,
  linkcolor=blue, citecolor=blue, urlcolor=blue
}

\begin{document}


\title{
Confinement transition to gravitational waves in the one-flavor \texorpdfstring{$SU(4)$}{SU(4)} Hyper Stealth Dark Matter theory
}

\author{V.~Ayyar}
\affiliation{Department of Physics, Boston University, Boston, MA 02215, USA}
\author{R.~C.~Brower}
\affiliation{Department of Physics, Boston University, Boston, MA 02215, USA}
\author{G.~T.~Fleming}
\affiliation{Theoretical Physics Division, Fermilab, Batavia, IL 60510, USA}
\author{J.~Ingoldby}
\affiliation{Institute for Particle Physics Phenomenology, Durham University, Durham DH1 3LE, UK}
\author{X.~Y.~Jin}
\affiliation{Computational Science Division, Argonne National Laboratory, Argonne, IL 60439, USA}
\author{N.~Matsumoto}
\email{nmatsum@bu.edu}
\affiliation{Department of Physics, Boston University, Boston, MA 02215, USA}
\affiliation{RIKEN/BNL Research Center, Brookhaven National Laboratory, 
Upton, NY 11973, USA}
\author{A.~S.~Meyer}
\affiliation{Physical and Life Sciences Division, Lawrence Livermore National Laboratory, Livermore, CA 94550, USA}
\affiliation{Nuclear Science Division, Lawrence Berkeley National Laboratory, Berkeley, CA 94720, USA}
\author{E.~T.~Neil}
\affiliation{Department of Physics, University of Colorado, Boulder, CO 80309, USA}
\author{J.~C.~Osborn}
\affiliation{Computational Science Division, Argonne National Laboratory, Argonne, IL 60439, USA}
\author{S.~Park}
\affiliation{Physical and Life Sciences Division, Lawrence Livermore National Laboratory, Livermore, CA 94550, USA}
\affiliation{Nuclear Science Division, Lawrence Berkeley National Laboratory, Berkeley, CA 94720, USA}
\affiliation{Department of Physics, Sejong University, Seoul 05006, South Korea}
\author{C.~T.~Peterson}
\affiliation{Department of Computational Mathematics, Science and Engineering, and Department of Physics and Astronomy, Michigan State University, East Lansing, MI 48824, USA}
\author{D.~Schaich}
\affiliation{Department of Mathematical Sciences, University of Liverpool, Liverpool L69 7ZL, UK}
\author{P.~Vranas}
\affiliation{Physical and Life Sciences Division, Lawrence Livermore National Laboratory, Livermore, CA 94550, USA}
\affiliation{Nuclear Science Division, Lawrence Berkeley National Laboratory, Berkeley, CA 94720, USA}
\author{O.~Witzel}
\affiliation{Center for Particle Physics Siegen (CPPS), Theoretische Physik 1, Naturwissenschaftlich-Technische Fakult\"at, Universit\"at Siegen, 57068 Siegen, Germany}

\collaboration{Lattice Strong Dynamics Collaboration}
\preprint{FERMILAB-PUB-26-0104-T, IPPP/26/17, SI-HEP-2026-02}

\date{\today}
\begin{abstract}
The thermodynamics of the $SU(4)$ gauge theory with a single flavor of fundamental quarks is analyzed on the lattice with dynamical fermion simulations, which is the low-energy sector of a realistic, strongly-interacting dark matter model---the Hyper Stealth Dark Matter.
The gravitational wave spectrum from the first-order confinement transition in the early universe is further calculated, where the effect of the dark sea quarks, which decrease the interface tension in the effective potential of the Polyakov loop, is shown numerically to lower the gravitational wave amplitude.
\end{abstract}
\maketitle
%

{\it Introduction---}The relevance of cosmology and astrophysics to particle physics is becoming increasingly profound, as they provide phenomenological grounds for the beyond-Standard-Model physics such as dark matter.
The purpose of this letter is to investigate the possibility that dark matter is a composite particle and its consequences in the early universe.
The Hyper Stealth Dark Matter (HSDM) model \cite{Fleming:2024flc}, pushing forward the idea of the Stealth Dark Matter (SDM) \cite{Appelquist:2015yfa,Appelquist:2015zfa}, embodies this possibility with an economical dark sector of single-flavor fundamental dark quarks and $SU(N_D)$ ($N_D\geq 3$) dark gluons, accompanied by the equilibration sector of heavy fermions with the full-strength electroweak interaction.

In the following, we focus on the dark sector and set $N_D=4$, which is the smallest $N_D$ for which the dark-matter-candidate baryon has bosonic statistics.
The generic phase structure of the one-flavor theory \cite{Fukugita:1987mb,Fukugita:1988qs,Alexandrou:1998wv,Shoji:2004yn,Takaishi:2007ga} can be found in the Columbia plot \cite{Brown:1990ev}:
The theory exhibits a first-order phase transition at a finite temperature $T_c$ for heavy quark mass starting from the quenched ({\it i.e.}, pure glue) limit, and as we lower the quark mass $m_q$, the first-order transition terminates at a critical mass $m_{q,c}$ (see also Ref.~\cite{Ayyar:2018ppa} for a study with multiple fermion representations).
The HSDM allows the baryon mass to be as light as about 50 GeV in the heavy-dark-quark limit.
The existence of the first-order phase transition raises further interest as a source of gravitational waves \cite{Caprini:2019egz,LISACosmologyWorkingGroup:2022jok}, whose detection will be supporting evidence for the strongly-interacting dark sector (see Ref.~\cite{Escriva:2024ivo} for discussion on crossovers). 

After the seminal work of Ref.~\cite{Schwaller:2015tja} that studied the gravitational waves from a wide class of composite dark sectors, Refs.~\cite{Huang:2020crf} and \cite{Pasechnik:2023hwv} reported quantitative predictions from the confinement transition of the pure glue theory with different numbers of colors and from the chiral transition with the linear sigma model, respectively, importing results from lattice calculations.
To some surprise, the results fell below the threshold of the near-future detectors.
On this ground, it is, of course, important to verify the result with independent calculations, and, at the same time, explore how the spectrum varies for different strongly coupled theories and understand the underlying physics, which help to develop accurate methodology \cite{Morgante:2022zvc,Bruno:2024dha,Huber:2025qbl,Houtz:2025ogg}.
State-of-the-art lattice calculations will play an indispensable role by providing a first-principles framework.

In this letter, we report on the lattice calculation of the one-flavor $SU(4)$ thermodynamics for the HSDM and a prediction of the gravitational wave spectrum (see Ref.~\cite{LatticeStrongDynamics:2020jwi} for related work for the SDM and Refs.~\cite{Bennett:2024bhy,Bennett:2025whm} for the pure glue $Sp(4)$ theory).
Concerning the theory space of composite dark matter in general, this amounts to adding a dark-quark-mass parameter to the $SU(N_D)$ Yang-Mills theory.
Since the anomaly takes an important role in the dynamics at least in the massless limit \cite{Leutwyler:1992yt,Creutz:2006ts}, we preserve the chiral structure by using domain-wall fermions \cite{Kaplan:1992bt,Shamir:1993zy,Furman:1994ky,Neuberger:1997bg} in the M\"obius formulation \cite{Brower:2012vk}.
Our analysis makes full use of the effective potential $V$ of the Polyakov loop \cite{Iwasaki:1993qu,Beinlich:1996xg,Wingate:2000bb,Smith:2013msa,Saito:2013vja,Iwami:2015eba}, whose shape is difficult to predict a priori in the presence of sea quarks \cite{Pisarski:2006hz,Pisarski:2016ixt} but is accessible using the histogram method \cite{Binder:1981sa} non-perturbatively on the lattice (see also Ref.~\cite{Moore:2000jw}); our estimate of $V$ does not rely on the ansatz for its functional form (see Ref.~\cite{Houtz:2025ogg} for extensions to fermionic theories).
With the obtained shape for $V$, the spectrum is calculated by following the outline of Ref.~\cite{Huang:2020crf}.
Preliminary results were reported in Refs.~\cite{Ayyar:2024dmt,Ayyar:2025eim}.


{\it Gravitational wave parameters}---Among several possible sources of gravitational waves \cite{Caprini:2015zlo}, we concentrate on the sound wave contribution.
Bubbles of the true vacuum nucleate during a supercooling process, which expand and collide to take over the universe eventually.
The temperature $T_*$ at which this transition happens is distinguished from the critical temperature $T_c$ of the first-order phase transition ($T_*<T_c$).

To make a prediction on the gravitational waves, following Refs.~\cite{Huang:2020crf,Morgante:2022zvc,Pasechnik:2023hwv,Huber:2025qbl}, we adopt an analysis with the effective potential $V$ by reducing the system into an effective theory of scalar variables \cite{Enqvist:1991xw,Kamionkowski:1993fg}, as assumed in the numerical simulations for the spectrum estimation \cite{Hindmarsh:2013xza,Hindmarsh:2015qta,Hindmarsh:2017gnf,Cutting:2018tjt}.
The scalar variable will be chosen as the Polyakov loop in the subsequent section.
We discuss the detectability in terms of the peak amplitude $h^2\Omega_{\rm peak}$ and frequency $f_{\rm peak}$, conventionally parametrized as \cite{Kamionkowski:1993fg, Huber:2008hg, Huang:2020crf,Pasechnik:2023hwv,Huber:2025qbl} (see Ref.~\cite{Caprini:2015zlo} for a review)
\begin{align}
    &h^2\Omega_{\rm peak}
    =
    2.65 \times 10^{-6}
    \,
    \Big(
    \frac{v_{w}}{\tilde\beta}
    \Big)
    \Big(
    \frac{
    \lambda_\alpha
    \kappa
    \alpha}{1+\alpha}
    \Big)^2
    \Big(
    \frac{100}{g_{\rm eff}}
    \Big)^{\frac{1}{3}},
    \label{eq:Omega_peak}
    \\
    &f_{\rm peak}
    =
    1.9\times10^{-5} \textrm{[Hz]}
    \,
    \Big(\frac{g_{\rm eff}}{100}\Big)^{\frac{1}{6}}
    \Big(
    \frac{T_*}{100[{\rm GeV}]}
    \Big)
    \Big(
    \frac{\tilde \beta}{v_{w}}
    \Big).
    \label{eq:fpeak}
\end{align}
While the parameters will be described in order below, the main input parameters turn out to be $\alpha$, $\tilde \beta$, and $T_*$.

For the wall velocity $v_w$, we use the Chapman-Jouguet detonation velocity \cite{Steinhardt:1981ct}:
\begin{align}
    v_J \equiv \frac{
    \sqrt{ \alpha(2 + 3\alpha) }
    + 1
    }
    {\sqrt{3}(1+\alpha)},
\end{align}
which serves as an optimistic estimate; see Ref.~\cite{Espinosa:2010hh}.
The efficiency factor $\kappa$ describes the kinetic energy fraction in the plasma fluid that produces the gravitational waves~\cite{Caprini:2019egz}, and is a complicated function of $\alpha$ and $v_w$ (see End Matter).
$g_{\rm eff} \equiv g_{{\rm eff},d}+g_{\rm eff, SM}$ is the total effective radiation degrees of freedom, and $g_{{\rm eff},d}$, $g_{\rm eff, SM}$ those from the dark and Standard Model sectors, respectively.
For the $N_f$-flavor $SU(N_D)$ theory, $g_{{\rm eff},d} = 2(N_D^2-1)+(7/8)4N_f$, assuming the deconfined phase. 
The temperature dependence of $g_{\rm eff,{\rm SM}}$ has a finite effect in our energy scale.
The factor $\lambda_\alpha \equiv g_{{\rm eff},d}/g_{\rm eff}$ takes into account the effect of visible matter on the spectrum.

The parameter $\alpha$ parametrizes the strength of the first-order transition \cite{Kamionkowski:1993fg,Espinosa:2010hh}, defined as the latent heat normalized by the effective degrees of freedom.
Following Refs.~\cite{Huang:2020crf,Morgante:2022zvc,Pasechnik:2023hwv,Huber:2025qbl}, we use the convenient expression in terms of the effective potential \cite{Hindmarsh:2017gnf}:
\begin{align}
    \alpha
     =
    -\frac{1}{3} 
    \Big(4 \Delta V - \frac{\partial \Delta V}{\partial \log T}\Big)/
    \frac{\partial \Delta V}{\partial \log T}
    \label{eq:alpha},
\end{align}
where $\Delta V \equiv V_+-V_-$ and $V_\pm$ are the values of $V$ in the deconfined ($+$) and confined ($-$) phases.
The decay rate $\Gamma$ is evaluated with the bounce action $S_3$ from the semiclassical analysis of $V$ as \cite{Coleman:1977py,Callan:1977pt}
\begin{align}
    \frac{\Gamma}{T^4} = \Big(\frac{S_3}{2\pi T}\Big)^{3/2} e^{-S_3/T},
    \label{eq:decay_rate}
\end{align}
by which the dimensionless decay constant is defined as 
\begin{align}
    \tilde \beta 
    \equiv -\frac{1}{H}\frac{d}{dt} \frac{S_3}{T}
    =
    \frac{d}{d\log T} \frac{S_3}{T}
    \label{eq:beta_tilde}
    ,
\end{align}
where $t$ is the cosmic time and $H$ the Hubble parameter.

As for the transition temperature $T_*$, following Refs.~\cite{Huang:2020crf,Pasechnik:2023hwv}, we use the percolation temperature $T_p$ defined by the relation $I(T_p)= 0.34$ \cite{Ellis:2018mja}, where
\begin{align}
    I(T)
    &\equiv
    \int_{T}^{T_c} \frac{dT_1}{H(T_1)}
    \frac{\Gamma(T_1)}{T_1^4}
    \frac{4\pi}{3}
    \Big[
    \int_T^{T_1} \frac{dT_2}{H(T_2)}
    v_w(T_2)
    \Big]^3,
    \label{eq:IT}
\end{align}
which depicts the total volume of the bubbles of the true vacuum.
The Friedmann equation expresses
\begin{align}
    H^2 = \frac{\rho_R}{3 M_{\rm pl}^2}, 
    \label{eq:Hubble}
\end{align} 
where $M_{\rm pl}=2.435 \times 10^{18}\, [{\rm GeV}]$ is the reduced Planck mass and $\rho_R = (\pi^2/30) g_{\rm eff} T^4$ the total radiation density.

{\it Polyakov loop effective potentials}---As is well known, the insertion of the Polyakov loop $L(\bx)$ can be interpreted as setting a probe quark at the spatial point $\textbf{x}$ \cite{Polyakov:1980ca}.
Upon proper renormalization, the expectation value measures the relative free energy of the static quark:
\begin{align}
    \langle L(\bx) \rangle
    \equiv e^{-(\Omega/T) (f_q - f)}
    \quad
    \big(
    Z \equiv e^{-(\Omega/T) f}
    \big),
    \label{eq:free_energy_L}
\end{align}
where $\Omega\equiv \int d^3\bx$ is the three-dimensional volume and $Z$ the partition function.

Though $\langle L(\bx) \rangle$ ceases to be an exact order parameter in the presence of sea quarks, when the confined and deconfined phases are separated by a first-order phase transition, we can expect a finite jump in $\langle L(\bx) \rangle$, denoting the discontinuous change in the required free energy to place a probe quark.
This makes the Polyakov loop a suitable observable for our study of thermodynamics.
The dimensionally-reduced effective action for the Polyakov loop may be written as
\begin{align}
    S_3
    &\equiv
    \int 
    d^3\bx\,
    \big(
    T^2 
    \vert \partial_i L \vert^2 
    + 
    V(L)
    \big),
\end{align}
where the temperature factor is supplied in the kinetic term for dimensionality \cite{Pisarski:2000eq}.
The potential $V$ is expected to have a double-well shape in the first-order regime.

To estimate $V$, we introduce the constrained effective potential $\bar{V}$ for the averaged Polyakov loop \cite{Fukuda:1974ey,ORaifeartaigh:1986axd,Wetterich:1991be,Bhattacharya:1992qb,KorthalsAltes:1993ca}, which is related to the normalized histogram $h(\bar{L})$ as
\begin{align}
    h(\bar{L}) 
    \equiv
    e^{-(\Omega/T) \, \bar{V}(\bar L)}
    &\equiv
    \Big\langle 
    \delta^2
    \Big(
    \bar L - \frac{1}{\Omega}\int d^3\bx
    \, L(\bx)
    \Big)
    \Big\rangle,
    \label{eq:def_histogram}
\end{align}
where $\delta^2(\bar L) \equiv \delta({\rm Re}\bar L)\delta({\rm Im}\bar L)$.
We immediately see that
\begin{align}
\int d^2\bar{L}\, 
     e^{-(\Omega/T) \, \bar{V}(\bar L)}
     \bar{L}
     =
     e^{-(\Omega/T) (f_q-f)},
     \label{eq:lint_L}
\end{align}
where $d^2\bar L \equiv d{\rm Re}\bar L \, d{\rm Im}\bar L$.
In the thermodynamic limit, the saddle point analysis becomes exact, and the location of the minimum $\bar L_0$ describes the expectation value of the averaged Polyakov loop ${\bar L}$.

Strictly speaking, $V$ and $\bar{V}$ must be distinguished as $\bar{V}$ becomes the convex hull of $V$ in the thermodynamic limit \cite{Heller:1983xg,ORAIFEARTAIGH1986653}. 
The potential barrier exists due to the interface tension \cite{Munster:1989we,Munster:1990yg,Bhattacharya:1992qb,Iwasaki:1993qu,Beinlich:1996xg,Giovannangeli:2002uv,Giovannangeli:2004sg,Pisarski:2006hz} in the mixed phase configuration ({\it i.e.}, between the local minima), which will be dominated in the infinite volume by the bulk free energy.
However, as was reported in Ref.~\cite{Heller:1983xg}, which is also the case in our study, the disappearance of the potential barrier turns out to be slower than the convergence of the tail part of the potential, and we find that $\bar V$ at a reasonably large but finite volume is a good estimate of $V$.
Though further volume scaling study of $\bar{V}$ \cite{Binder:1981sa,Brezin:1985xx,Iwasaki:1993qu} would be beneficial, this is beyond the scope of this paper.

{\it Renormalization}---The continuum dimensionful quantities estimated from the lattice are dimensionally transmuted and measured at a reference mass scale $\mu$, which we set to the Wilson flow scale \cite{Luscher:2010iy}: $\mu\equiv1/\sqrt{t_0}$ (see End Matter for more detail). 
The value of $1/\sqrt{t_0}$ in GeV defines the overall scale of the dark sector.
We define the line of constant physics with the mass $M_B \sqrt{t_0}$ of the lightest baryon.
$M_B \sqrt{t_0}$ and $1/\sqrt{t_0}$ in GeV are thus the inputs to the theory.

To renormalize the Polyakov loop \cite{Polyakov:1980ca,Kaczmarek:2002mc}, the Wilson flow \cite{Petreczky:2015yta} can be utilized most conveniently, where different flow times correspond to different renormalization schemes.
We measure the renormalized Polyakov loop $L(\bx)$ at the theory scale $1/\sqrt{t_0}$.

{\it Lattice setup}---We use M\"obius domain-wall fermions \cite{Brower:2012vk}
with the Wilson gauge action \cite{Wilson:1974sk}
on lattices of size
$N_s^3 \times N_\tau$.
The auxiliary dimension $L_s=8$ for the domain-wall fermion is set such that the residual mass is $O(10^{-4})$ \cite{Ayyar:2025eim}. 
The dynamical simulation is performed with the exact one-flavor hybrid Monte Carlo algorithm \cite{Chen:2014hyy} using the Grid software library \cite{Boyle:2015tjk}.

To study the thermodynamics, we fix $N_\tau=8$ and vary the gauge coupling $\beta \equiv 2N_D/g^2$ for the dark-quark mass $\hat{m}_q \equiv m_q a = 0.1, 0.2, 0.3, 0.4$.
Since the temperature is $T = 1/(N_\tau a)$ and the gauge coupling determines the lattice spacing $a$, varying $\beta$ amounts to varying $T$.
Finite volume effects are studied by taking $N_s=16,24,32$.
Zero-temperature ensembles are generated for the scale setting with $N_s=24$ and $N_\tau = 48$.
More details on lattice calculation are provided in the Supplementary Material.

Ideally, the continuum limit should be taken by varying $N_\tau$.
While discretization effects will be studied systematically in future work, 
$O(a^2)$ corrections with domain-wall fermions are expected to be subdominant to the relatively large statistical error in the final results.
See, e.g., the $a^2$-dependence of the baryon mass in End Matter.

\begin{figure}
    \centering
    \includegraphics[width=0.45\linewidth]{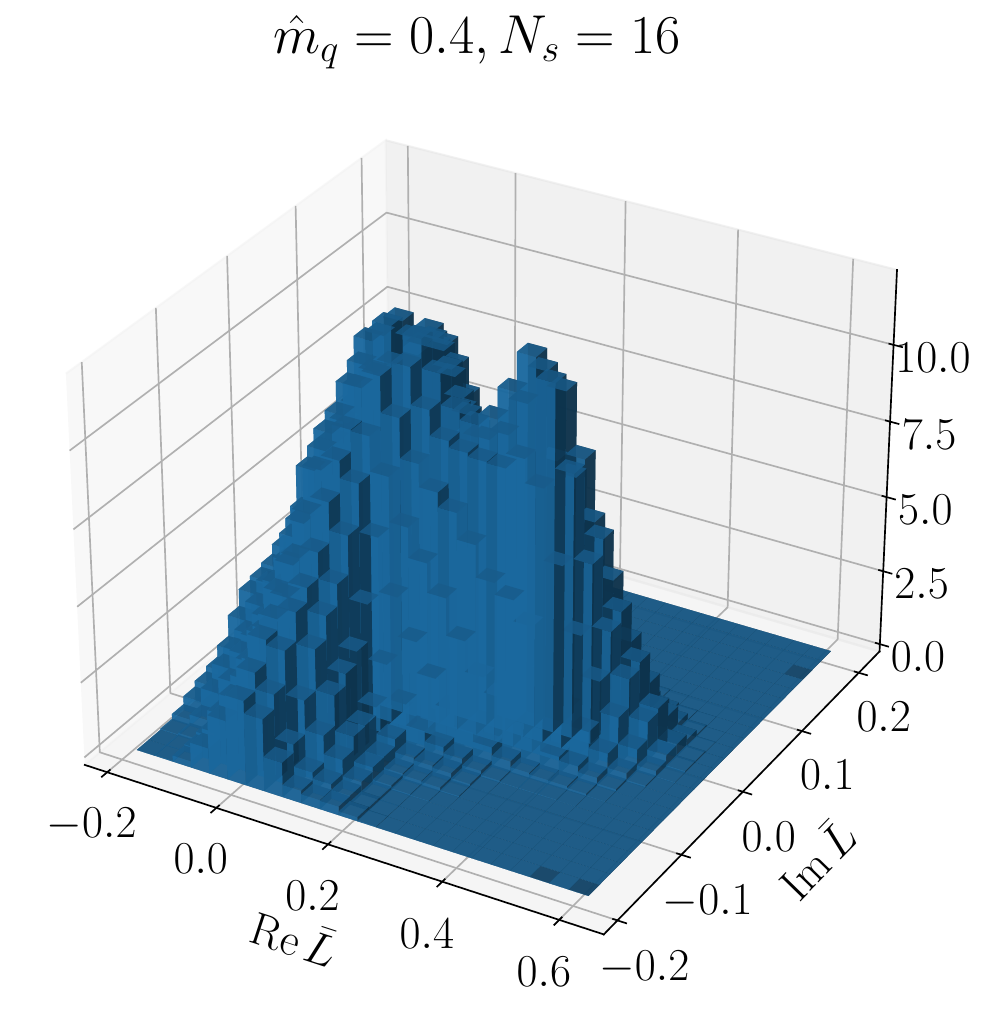}
    \includegraphics[width=0.45\linewidth]{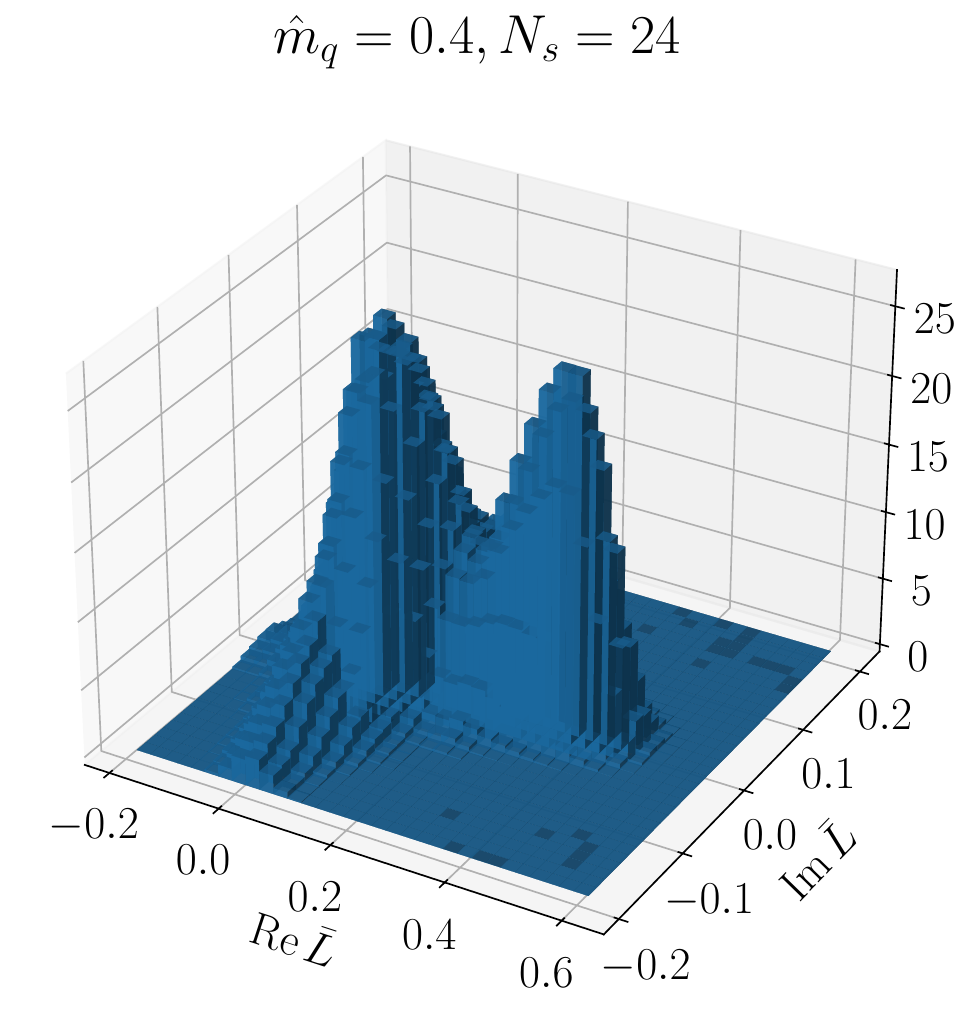}
    \includegraphics[width=0.45\linewidth]{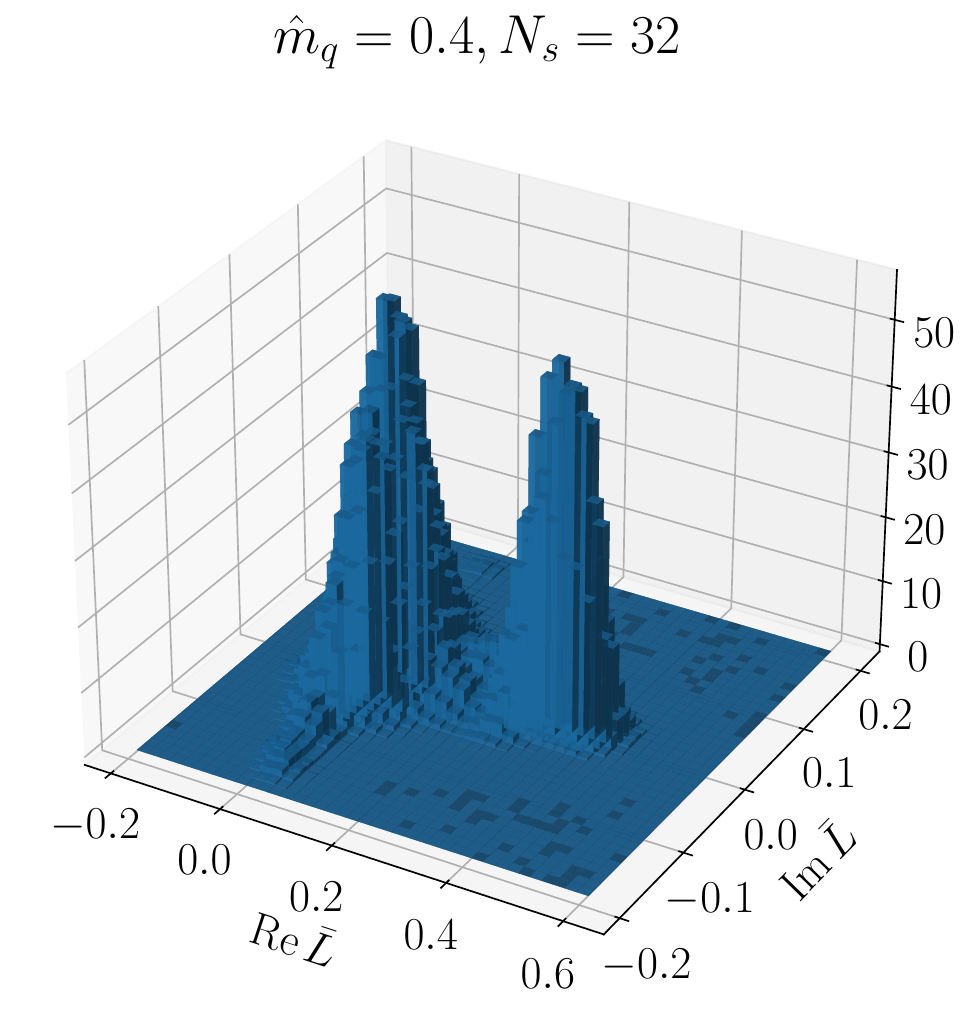}
    \caption{The histogram $h(\bar{L})$ for the averaged Polyakov loop $\bar{L}$ at $\beta_c$ with $\hat{m}_q=0.4$ and $N_s=16,24,32$ from top left to bottom. 
    Errorbar is suppressed. }
    \label{fig:histogram}
\end{figure}

\begin{figure}[hbt]
    \centering
    \includegraphics[width=0.45\linewidth]{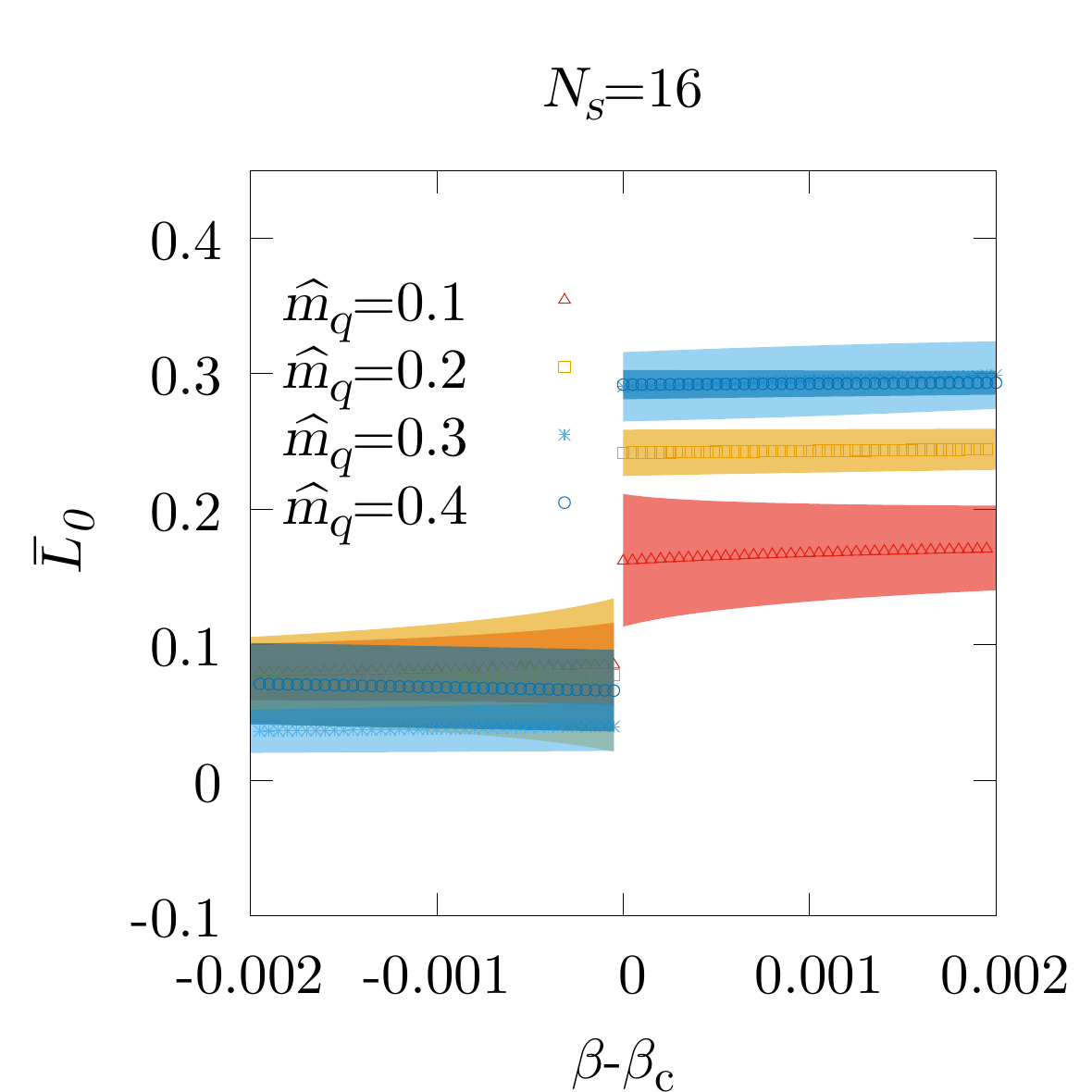}
    \includegraphics[width=0.45\linewidth]{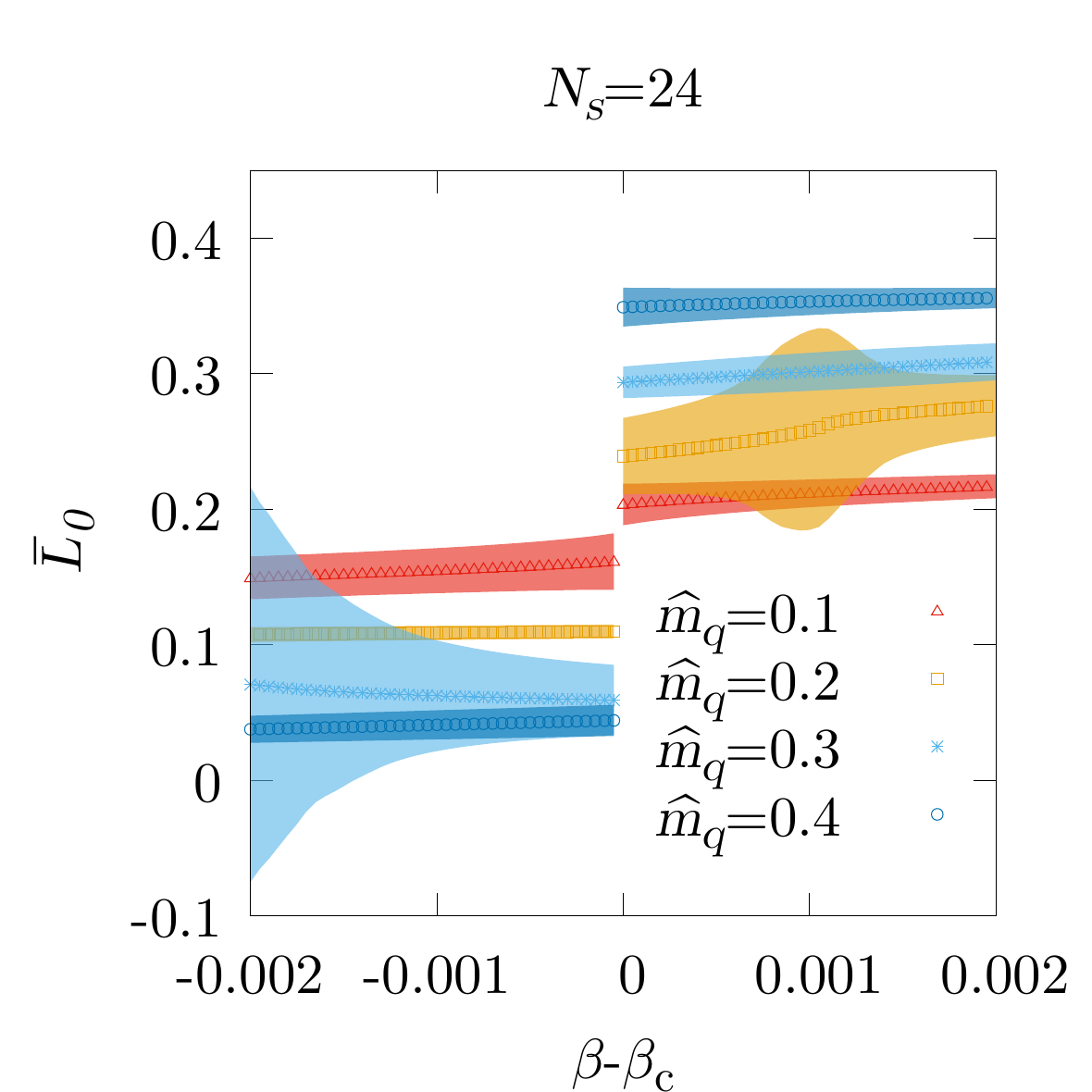}
    \includegraphics[width=0.45\linewidth]{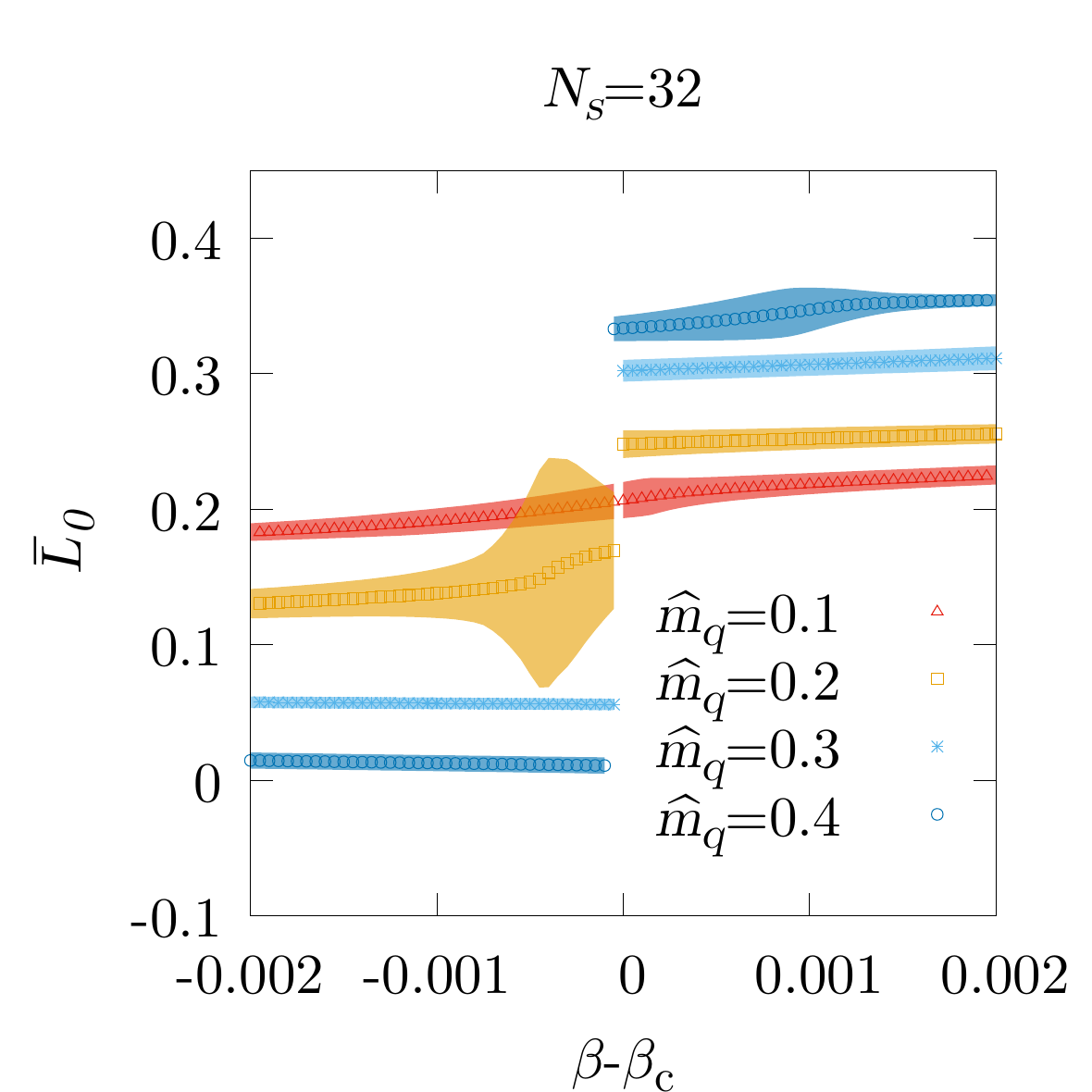}
    \caption{The location of the minimum $\bar{L}_0$ for $N_s=16,24,32$ from top left to bottom.
    $\hat{m}_q=0.1$ does not exhibit a discontinuous jump for $N_s=32$, implying a crossover, whose $\beta_c$ is  determined by the largest derivative $d\bar{L}_0/d\beta$. }
    \label{fig:L0jump}
\end{figure}

{\it Phase transition}---We evaluate the histogram~\eqref{eq:def_histogram} by regularizing the delta function with the Gaussian. 
By using the multipoint reweighting method \cite{Ferrenberg:1988yz, Ferrenberg:1989ui} following Refs.~\cite{Saito:2013vja,Iwami:2015eba}, we can obtain the histogram for an arbitrary value of $\beta$ around $\beta_c$ (see Fig.~\ref{fig:histogram}).
We use a cubic spline to interpolate the effective potential $\bar{V}$ on the real axis between the discrete points.
The order of the transition and the critical coupling $\beta_c$ are identified by looking for the discontinuity of the minimum $\bar{L}_0$ (see Fig.~\ref{fig:L0jump}), which is a robust characteristic of the first-order transition that exists even when $\bar{V}$ is the convex hull.
In the case of the $N_s=32$ result, the gap disappears for $\hat{m}_q=0.1$, implying a crossover; the apparent discontinuity in the smaller volumes may be attributed to the large correlation length around the second-order point. 
We identify the $\hat{m}_{q} \geq 0.2$ ensembles to be in the first-order regime.

\begin{figure}[thb]
    \centering
    \includegraphics[width=0.45\linewidth]{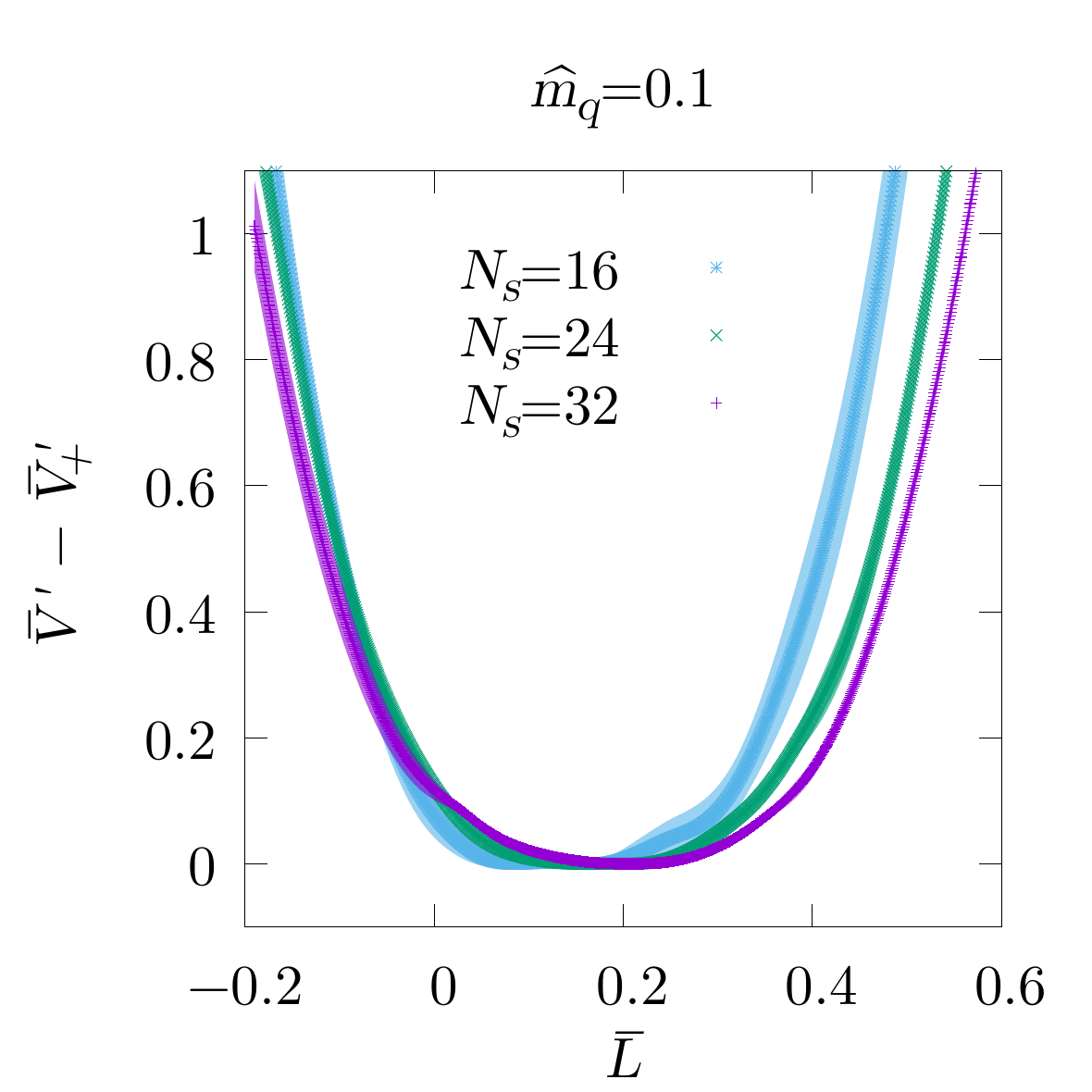}
    \includegraphics[width=0.45\linewidth]{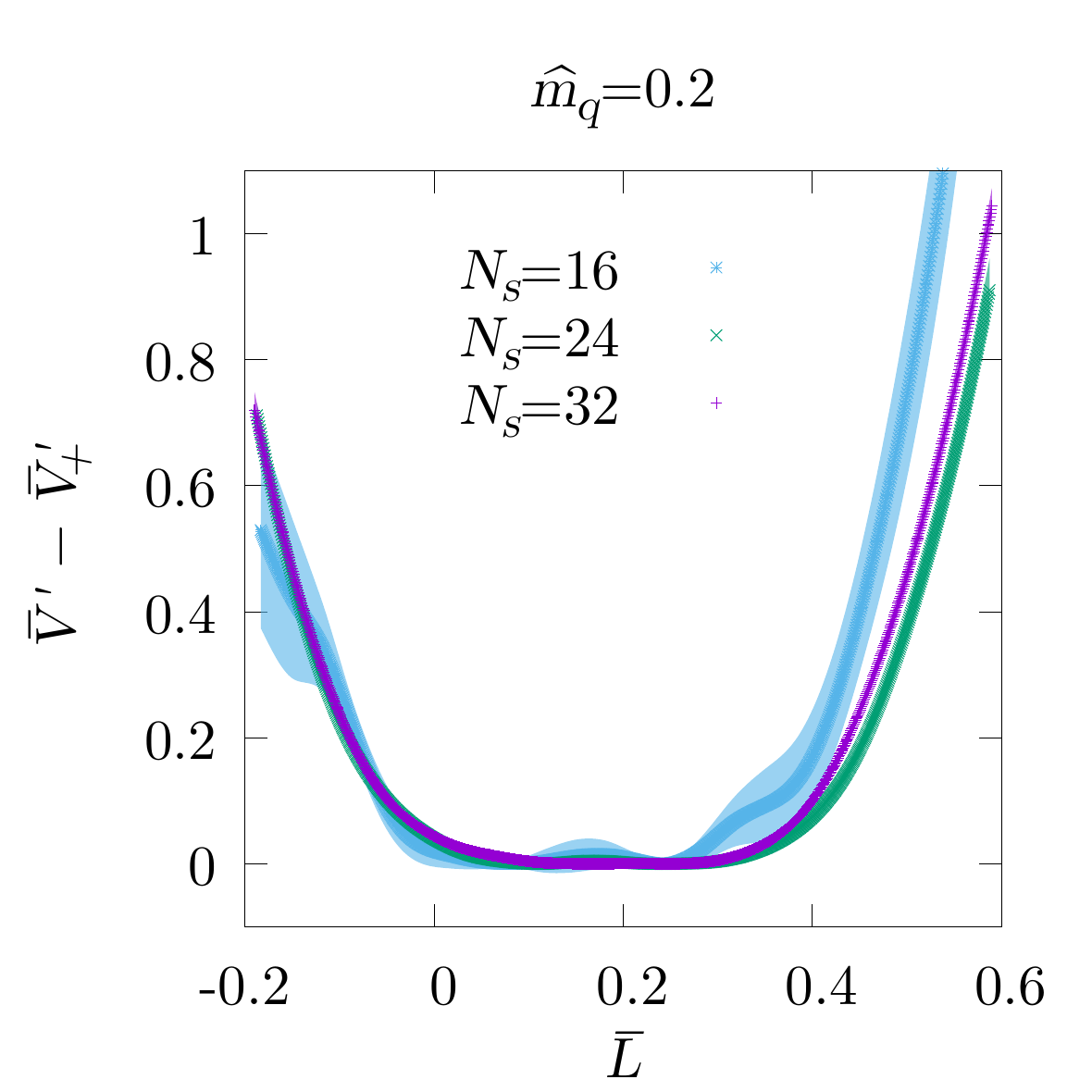}
    \includegraphics[width=0.45\linewidth]{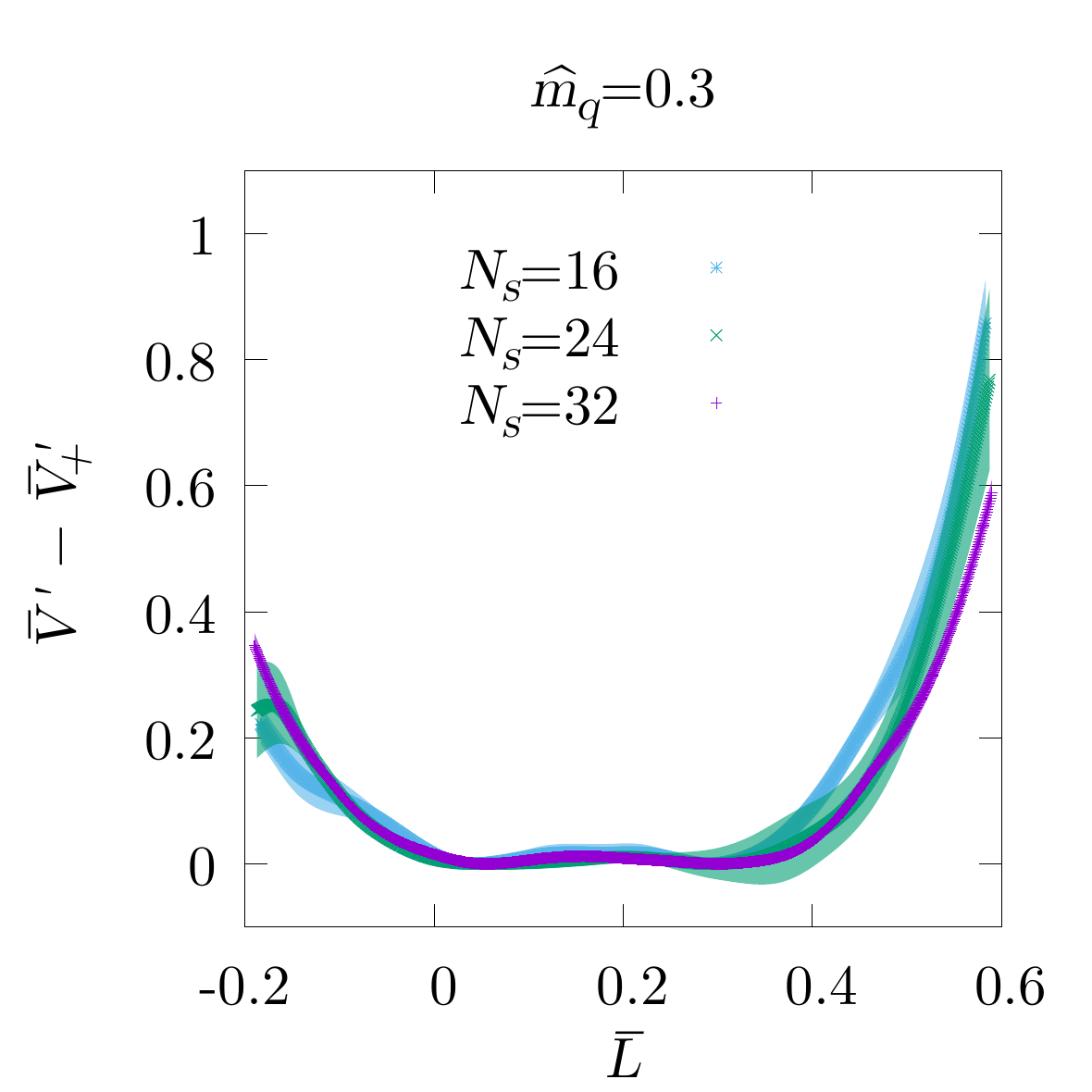}
    \includegraphics[width=0.45\linewidth]{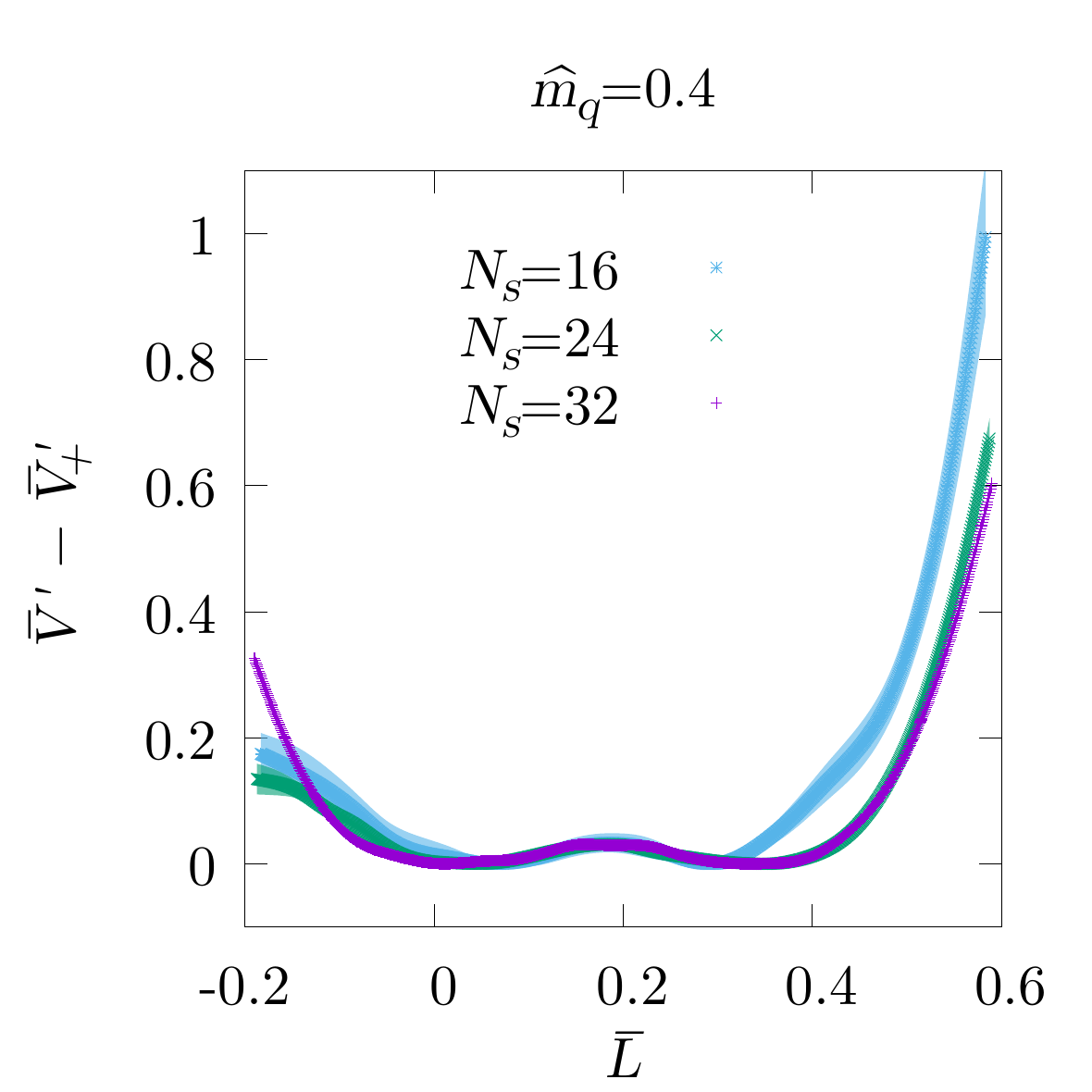}
    \caption{ The dimensionless constrained effective potential $\bar{V}'\equiv\bar{V}/T^4$ along the real axis at $\beta_c$. $\hat{m}_q=0.1,0.2,0.3,0.4$ from top left to bottom right. }
    \label{fig:DeltaV_c}
\end{figure}

Finite volume effects can be observed pictorially in the reconstructed potential at $\beta_c$ in Fig.~\ref{fig:DeltaV_c}.
The sizable effect for $N_s=16$ can be understood from the histogram on the complex $\bar{L}$ plane (Fig.~\ref{fig:histogram}) as the remnant of the $\mathbb{Z}_4$ symmetry in the pure glue theory; the soft breaking by the fermionic boundary condition takes effect in larger volumes. 
We also observe in Fig.~\ref{fig:DeltaV_c} that the lowering of the potential barrier is hardly visible for $N_s=24,32$.
We thus use the reconstructed $\bar{V}$ for $N_s=32$ as the estimate of the potential $V$.

\begin{figure}[htb]
    \centering
    \includegraphics[width=0.8\linewidth]{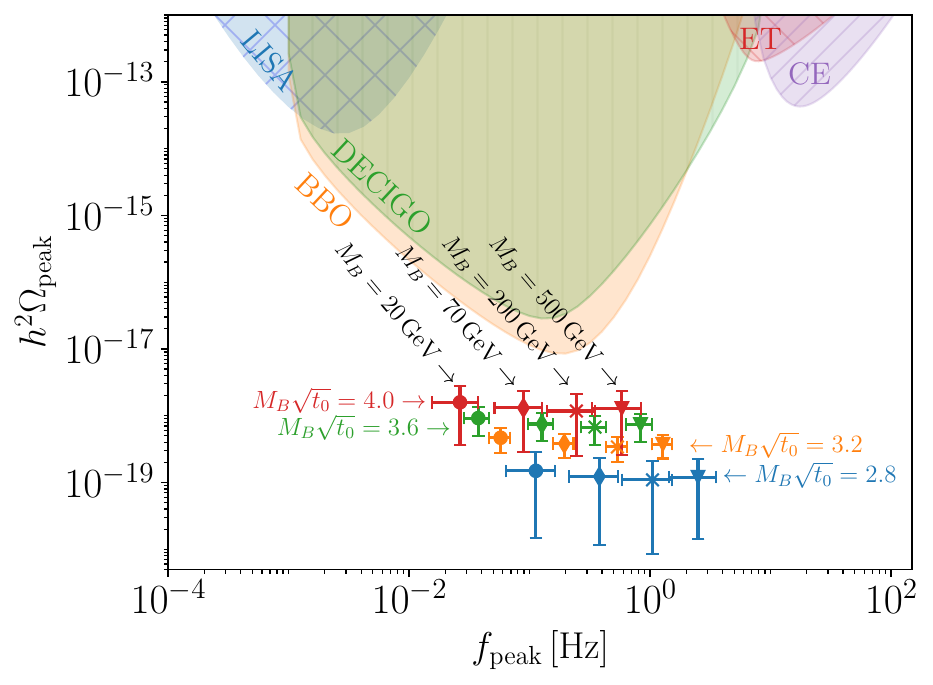}
    \caption{The peak amplitude $h^2\Omega_{\rm peak}$ and frequency $f_{\rm peak}$, compared against near-future detectors' sensitivity, taken from Ref.~\cite{Schmitz:2020syl}.
    }
    \label{fig:sensitivity}
\end{figure}

\begin{figure}[htb]
    \centering
    \includegraphics[width=0.48\linewidth]{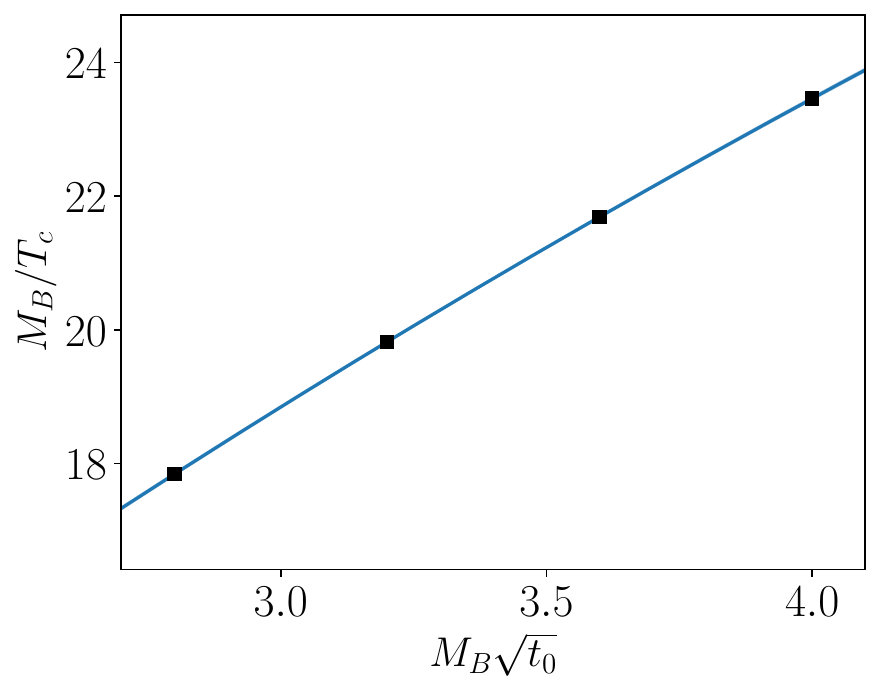}
    \includegraphics[width=0.48\linewidth]{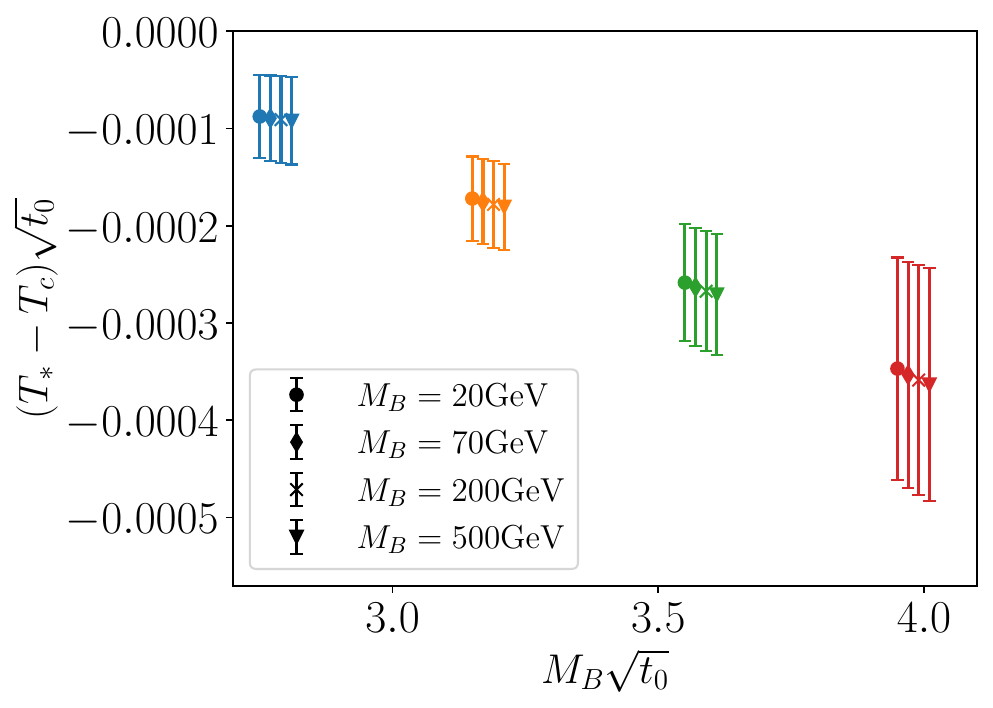}
    \caption{ (Left) $M_B/T_c$ and (Right) $(T_*-T_c)\sqrt{t_0}$ as functions of $M_B\sqrt{t_0}$.
    In the right panel, 
    points with equal $M_B\sqrt{t_0}$ are horizontally displaced for visibility,
    and colors and markers are varied as in Fig.~\ref{fig:sensitivity}.
    }
    \label{fig:MB/Tc}
\end{figure}

{\it Gravitational wave spectrum}---
From the estimate of the potential $V$, the bounce action $S_3$ is calculated with the shooting method.
Combined with the scale setting results, the data points are obtained as dimensionless tuples $(S_3/T, (T-T_c)\sqrt{t_0}, M_B\sqrt{t_0})$ that can be fitted to express $S_3/T$ as a continuous function of $(T-T_c)\sqrt{t_0}$ and $M_B\sqrt{t_0}$.
The dimensionless vacuum energy $\Delta V/T^4$ can be expressed similarly as a function of $(T-T_c)\sqrt{t_0}$ and $M_B\sqrt{t_0}$. 
See End Matter for more on the intermediate steps.
With the continuous functions, the gravitational wave parameters can be readily evaluated, as shown in Fig.~\ref{fig:sensitivity}.
Varying $M_B\sqrt{t_0}$ effectively changes the quark mass $\hat{m}_q$, whereas varying $M_B$ in GeV rescales the overall theory scale $1/\sqrt{t_0}$. 
The rescaling shifts the peak frequency while leaving the peak amplitude approximately unchanged, and therefore moves the points horizontally in Fig.~\ref{fig:sensitivity}.
As shown in Fig.~\ref{fig:MB/Tc}, $M_B/T_c \sim 20$ in our parameter range, and $-0.0005 < (T_*-T_c)\sqrt{t_0}$, where the interpolating functions are valid.

\begin{figure}[htb]
    \centering
    \includegraphics[width=0.48\linewidth]{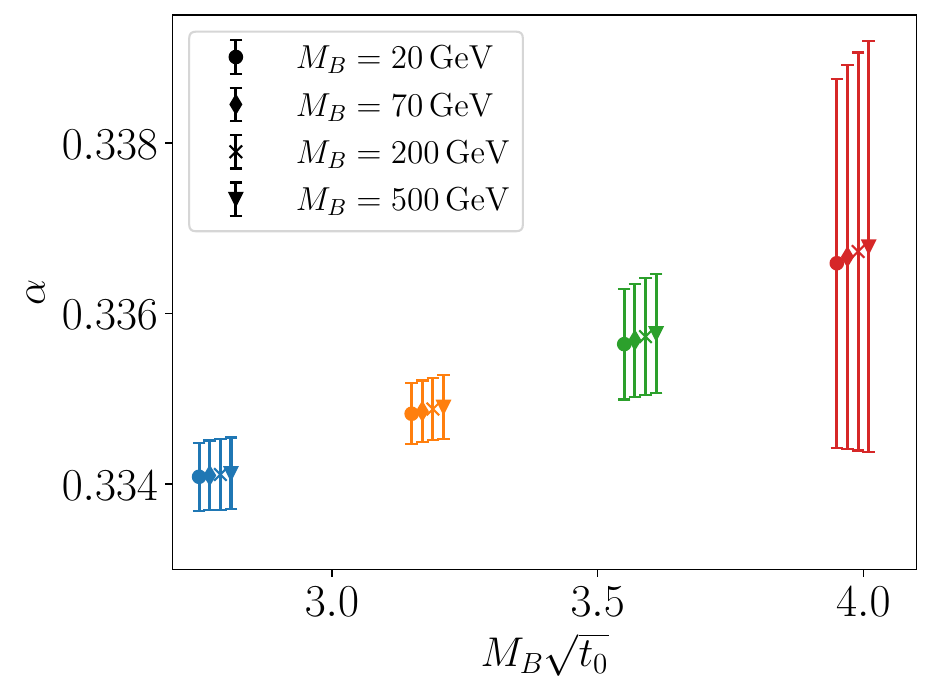}
    \includegraphics[width=0.48\linewidth]{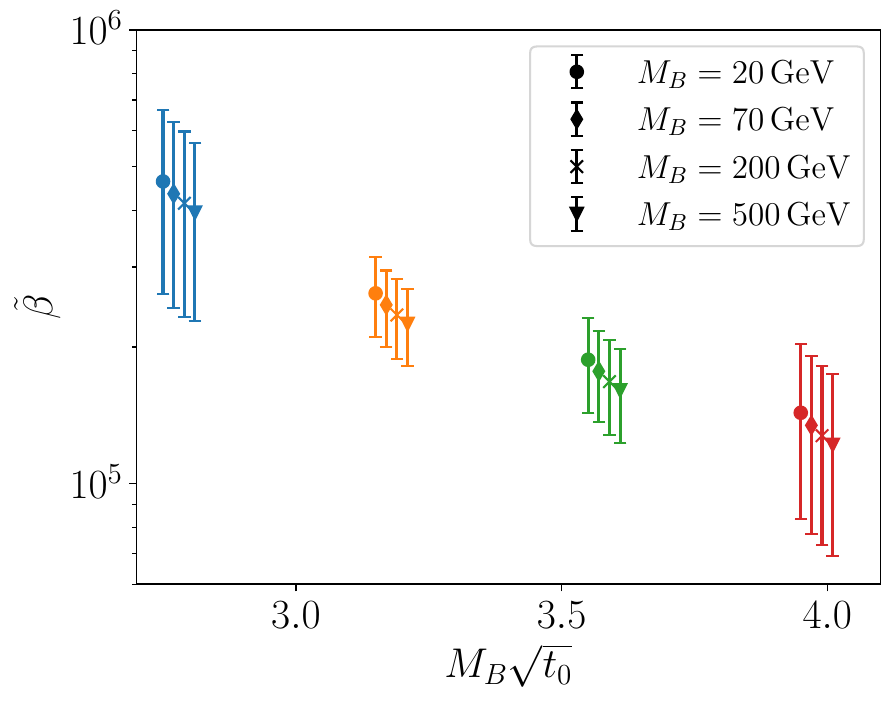}
    \caption{ (Left) The strength parameter $\alpha$ and (Right) the dimensionless decay constant $\tilde\beta$. 
    Points with equal $M_B\sqrt{t_0}$ are horizontally displaced for visibility.
    Minor dependences on $M_B$ [GeV] are from the comparison of $1/\sqrt{t_0}$ to the Hubble parameter $H$ and to the temperature $T$ in GeV that enters into $g_{\rm eff, SM}(T)$. 
    }
    \label{fig:alpha_beta}
\end{figure}

The $M_B \sqrt{t_0}$ dependence of the spectrum can be understood in terms of $\alpha$ and $\tilde\beta$ (Fig.~\ref{fig:alpha_beta}).
We see that the larger interface tension observed as the larger barrier height in Fig.~\ref{fig:DeltaV_c} gives a stronger transition and a longer lifetime of the bubbles, both of which increase the amplitude.

{\it Discussion}---The amplitude saturates as we reach the pure glue limit, $M_B\sqrt{t_0} \to \infty$.
Our result for the largest baryon mass $M_B\sqrt{t_0}=4.0$ may be considered as near the pure glue limit, and is in agreement with the former works on the pure glue theory \cite{Huang:2020crf,Morgante:2022zvc,Huber:2025qbl} 
in order of magnitude.
The amplitude may be amplified by varying the number of colors $N_D$ \cite{Schwaller:2015tja,Huang:2020crf,Huber:2025qbl} or the Lie group \cite{Bruno:2024dha}.

Note that the observed quark-mass dependence is mainly from the change in the decay rate:
In our application, we have $\Delta V \ll \partial \Delta V/\partial \log T$ in Eq.~\eqref{eq:alpha}, and the strength parameter is consistently $\alpha \sim 1/3$, as in Refs.~\cite{Huang:2020crf,Morgante:2022zvc,Huber:2025qbl} (see Fig.~\ref{fig:alpha_beta}).
It is further noteworthy that such dependence of $\alpha$ relies on the calculation scheme with the effective scalar variable.
To understand the sea-quark effects better, calculating gluonic and fermionic contributions to the latent heat \cite{Boyd:1996bx,Lucini:2005vg,Borsanyi:2010cj,Shirogane:2016zbf} separately may be useful.
In combination with the lattice methods for the decay rate \cite{Shen:2022xcm,Jin:2025nkw}, one can obtain $\alpha$ and $\tilde{\beta}$ without introducing the effective potential $V$.

Another simplification in the calculation is the wall velocity $v_w=v_J$ that assumes detonation.
Accurate determination of $v_w$ requires evaluation of hydrodynamic quantities such as fluid velocity in the two phases \cite{Espinosa:2010hh,Huber:2025qbl}.
It will be interesting to develop a lattice framework in parallel to holographic models \cite{Bea:2021zsu,Bigazzi:2021ucw,Janik:2022wsx,Chen:2022cgj}.

Concerning the HSDM as the theory of dark matter, glueballs and mesons, together with baryons, provide essential inputs to its phenomenology \cite{Fleming:2024flc}.
Detailed spectroscopy and scattering studies will follow in future work (see also Refs.~\cite{Farchioni:2007dw, Jaeger:2022ypq, DellaMorte:2023ylq, Bergner:2023oqs}).
Our spectroscopy result for the connected diagrams (see End Matter) suggests that, together with the glueball masses of the quenched theory \cite{Lucini:2010nv}, the critical mass $m_{q,c}$ roughly corresponds to where the mass hierarchy between glueball and mesonic states turns over.
The simple setup of the one-flavor theory also serves as a theoretically intriguing system for scrutinizing the confining dynamics of quantum chromodynamics.

{\it Acknowledgments}---We thank Hooman Davoudiasl, Anna Hasenfratz, Rob Pisarski, and Claudio Rebbi for valuable discussions. 
S.P., A.S.M., and P.M.V.\ would like to thank Scott Futral of LLNL for his support and early access to LLNL's exascale systems, Tuolumne and El Capitan.
This work was performed under the auspices of the U.S.\ Department of Energy (DOE) by Lawrence Livermore National Laboratory (LLNL) under Contract {DE-AC52-07NA27344}.
A.S.M.\ is supported in part by Neutrino Theory Network Program Grant {DE-AC02-07CHI11359}, and DOE Award {DE-SC0020250}.
S.P.\ acknowledges the support from the ASC COSMON project.
N.M. and C.P.\ are supported in part by the Scientific Discovery through Advanced Computing (SciDAC) program under FOA LAB-2580, funded by DOE, Office of Science.
N.M.\ is further supported in part by DOE under Award {DE-SC0015845}.
D.S.\ is supported by UK Research and Innovation Future Leader Fellowship {MR/X015157/1} as well as Science and Technology Facilities Council consolidated grant {ST/X000699/1}.
This document was prepared by the Lattice Strong Dynamics collaboration using the resources of the Fermi National Accelerator Laboratory (Fermilab), a U.S. Department of Energy, Office of Science, Office of High Energy Physics HEP User Facility. Fermilab is managed by Fermi Forward Discovery Group, LLC, acting under Contract No.~89243024CSC000002.
Results presented in this letter were produced using Grid \cite{Boyle:2015tjk}.



\bibliography{ref} 


\appendix*

\section{End Matter}

After introducing (i) the formulas for the efficiency factor $\kappa$, we elaborate on (ii) the scale setting, (iii) the bounce action, and (iv) the interpolating functions.
It is convenient to measure the dimensionful quantities related to the gravitational wave in the units of $T$ \cite{Huang:2020crf}, which will be indicated by the prime symbol.
Quantities measured in the units of $a^{-1}$ are indicated conventionally with the hat symbol.
We further work in the units where $1/\sqrt{t_0} = 1$ for notational simplicity.
(v) Partial spectroscopy results for baryonic and mesonic states are also given.

\subsection{(i) Efficiency factor}

Following Refs.~\cite{Huang:2020crf,Morgante:2022zvc,Pasechnik:2023hwv,Huber:2025qbl}, we use the expression:
\begin{align}
    \kappa \equiv 
    \sqrt{\tau_{\rm sw}} \kappa_{v}.
\end{align}
The factor $\sqrt{\tau_{\rm sw}}$ takes into account the lifetime of the sound-wave source \cite{Ellis:2020awk, Guo:2020grp}:
\begin{align}
    \tau_{\rm sw}
    &=
    1-
    \left[1+2\frac{(8\pi)^{\frac{1}{3}} v_{w}}
    {\tilde\beta U_f}
    \right]^{-1/2},
    \quad
    U_f^2
    =
    \frac{3}{4}
    \frac{\alpha}{1+\alpha}
    \kappa_{v},
\end{align}
which lowers the amplitude in numerics.
$\kappa_{v}$ is the factor that is derived from hydrodynamical analysis, and is given for the Jouguet detonation as \cite{Espinosa:2010hh}
\begin{align}
    \kappa_{v}(v_w=v_J)
    =
    \frac{\sqrt{\alpha}}{0.135+\sqrt{0.98+\alpha}}.
\end{align}

\subsection{(ii) Scale setting}

\begin{figure}[htb]
    \centering
    \includegraphics[width=0.55\linewidth]{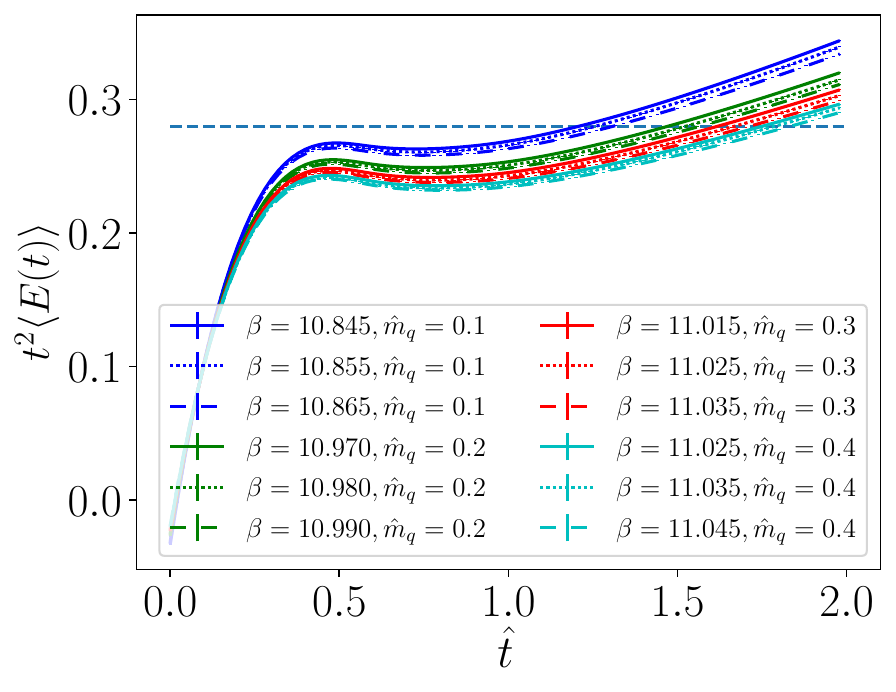}
    \caption{$t^2 \langle E(t) \rangle$ plotted against $\hat{t}$ for the zero-temperature ensembles.
    The dashed line indicates the defined scale $t^2 \langle E(t) \rangle|_{t=t_0}
    = 0.28$. }
    \label{fig:Eflow}
\end{figure}

Our definition of the Wilson flow scale $t_0$ is (see Fig.~\ref{fig:Eflow})
\begin{align}
    t^2 \langle E(t) \rangle|_{t=t_0}
    = 0.28,
    \label{eq:constraint}
\end{align}
where $E(t)$ is the flowed energy density at the flow time $t$ \cite{Luscher:2010iy}.
Since we also use $t_0$ to renormalize the Polyakov loop, the right-hand side is chosen such that $\sqrt{8 \hat{t}_0} < N_\tau/2$ for $N_\tau=8$, by which the flow does not wrap around the temporal extent in the finite-temperature calculations.
See Refs.~\cite{Hirakida:2018uoy,Butti:2025rnu} on the $N_D$ scaling from the conventional choice.

\begin{figure}
    \centering
    \includegraphics[width=0.55\linewidth]{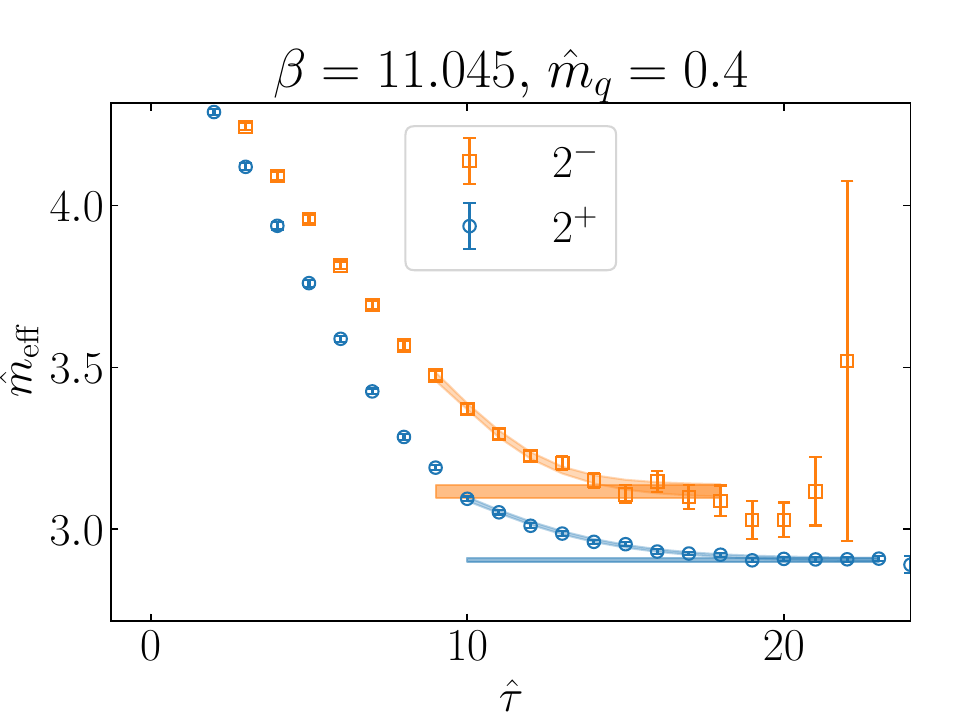}
    \caption{Effective masses $\hat{m}_{\rm eff}$ for the $J^P = 2^{+}$ and $2^{-}$ correlators with $\beta=11.045$ and $\hat{m}_q=0.4$, plotted against temporal slices $\hat{\tau}$.
    The reduced $\chi^2$ is 0.91 for $2^+$ and 0.57 for $2^-$.
    }
    \label{fig:Meff}
\end{figure}

For the one-flavor theory, local baryon operators have totally symmetric flavor indices and totally anti-symmetric spin indices.
Accordingly, the lowest baryon states are spin 2.
Figure~\ref{fig:Meff} shows the effective masses for the $J^P=2^{+},2^{-}$ point-to-point correlators, which are fitted by the ansatz: $\hat{M}_0 + A \exp(-\Delta \hat{M} \,\hat{\tau})$. 
The mass $\hat{M}_0$ of the $2^+$ state is used as $\hat{M}_B$ to define the line of constant physics.

\subsection{(iii) Bounce action}

\begin{figure}[htb]
    \centering
    \includegraphics[width=0.45\linewidth]{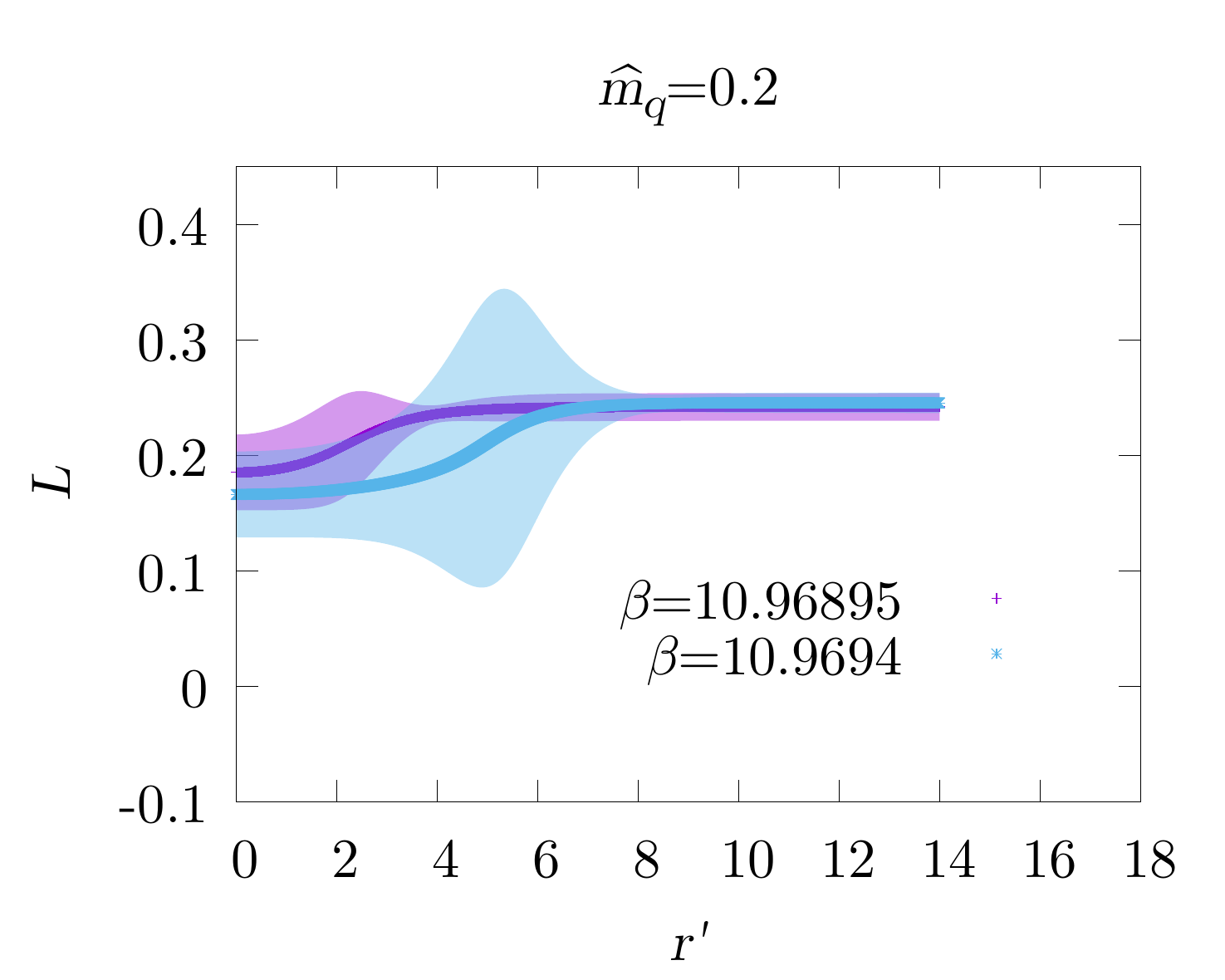}
    \includegraphics[width=0.45\linewidth]{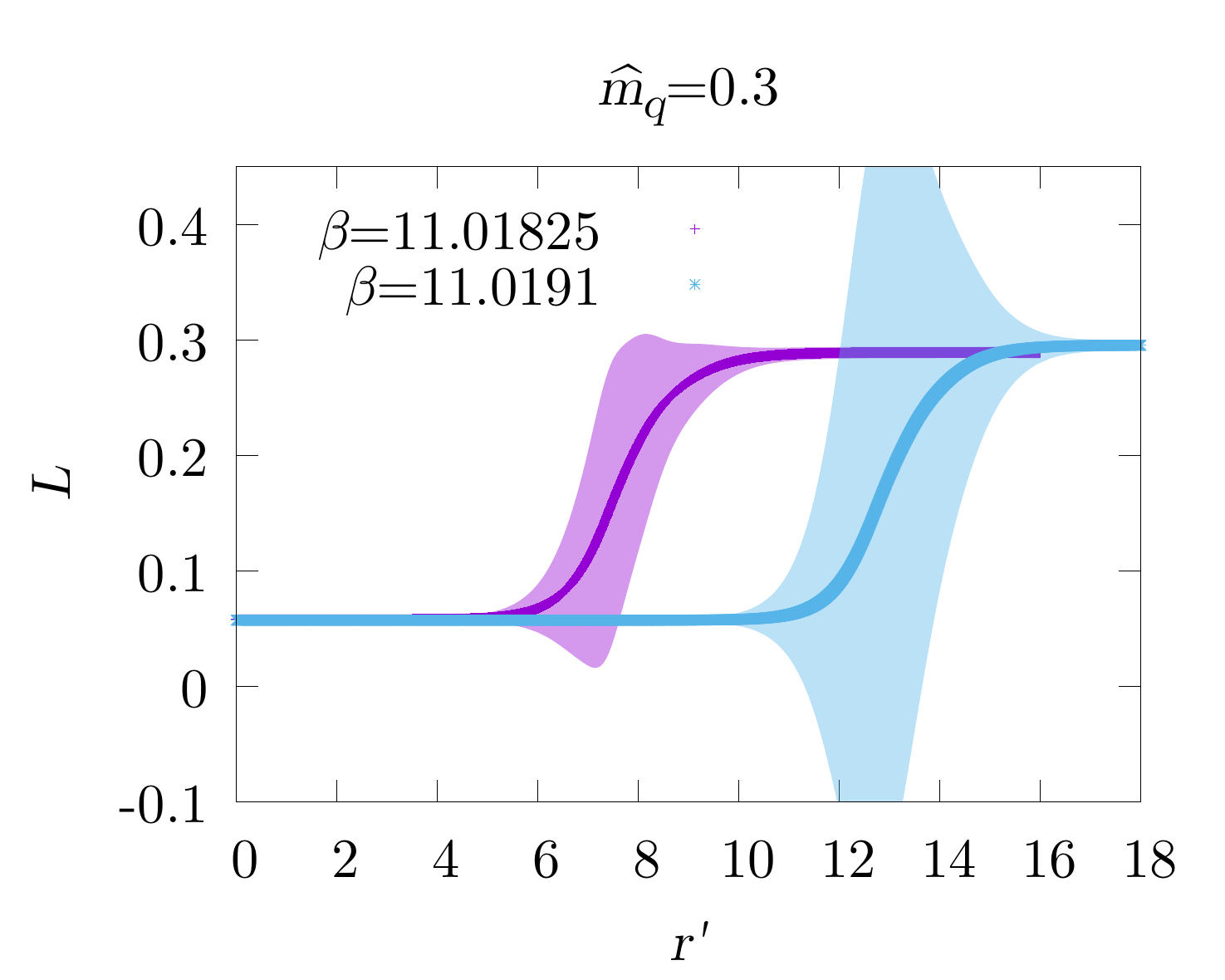}
    \includegraphics[width=0.45\linewidth]{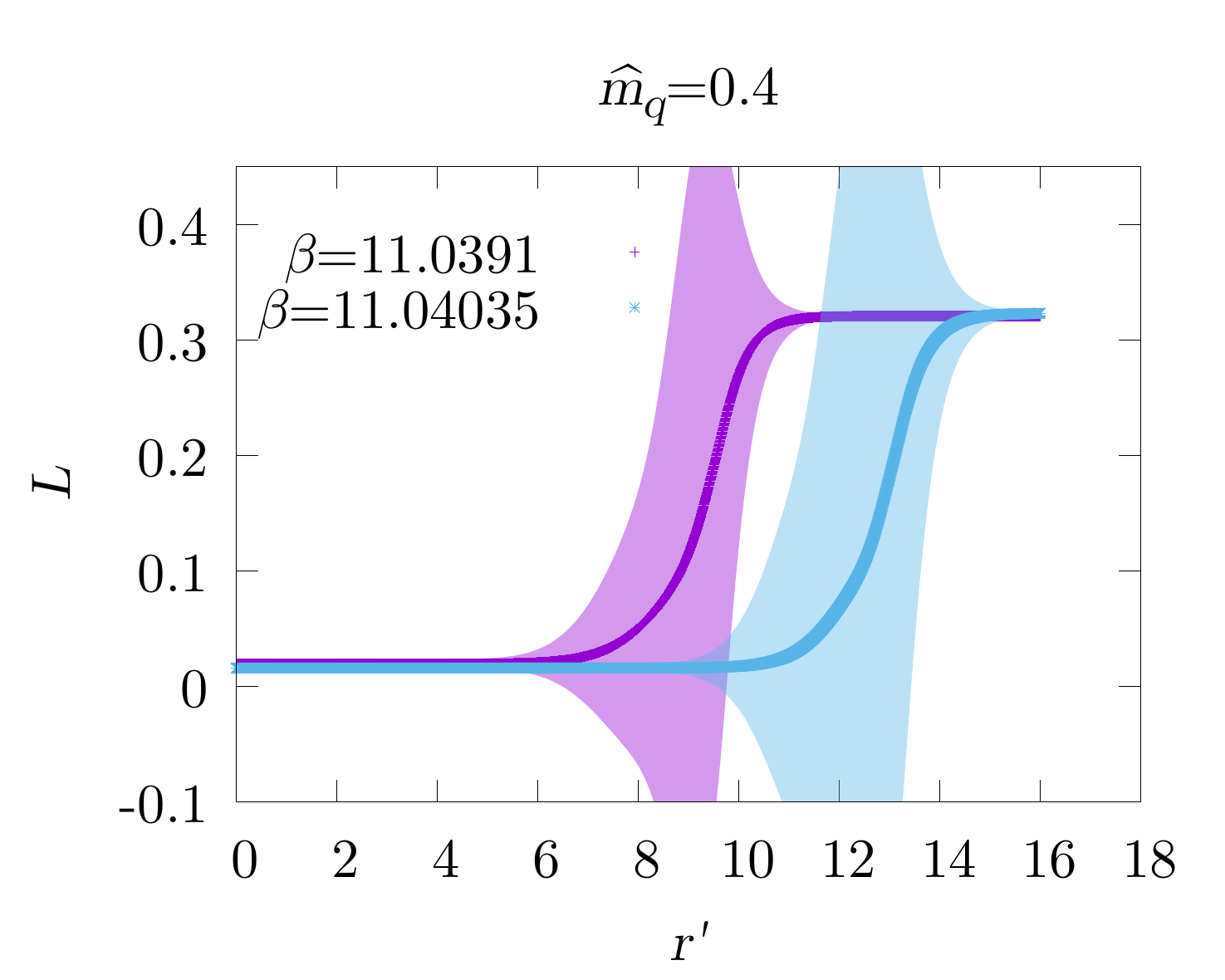}
    \caption{The bounce solution $L(r')$ for the smallest and largest $\beta$ in the fitted range for $S_3'$. $\hat{m}_q=0.2,0.3,0.4$ from top left to bottom.}
    \label{fig:bounce}
\end{figure}

To obtain the bounce solution \cite{Coleman:1977py},  the equation of motion \cite{Huang:2020crf}
\begin{align}
    -\frac{d^2 L}{dr^{\prime 2}}
    -\frac{2}{r'}\frac{d L}{dr'}
    +\frac{1}{2}\frac{d V'}{dr'} = 0
    \label{eq:EOM}
\end{align}
is solved with the boundary conditions $dL(r'=0)/dr' = 0$ and $L(r'=\infty) = L_+$ by the shooting method.
Obtained solutions are shown in Fig.~\ref{fig:bounce} for the smallest and largest $\beta$ which enter the fit of $S'_3$, where the largest value of $\beta$ is constrained by the machine precision in the shooting method.
By substituting the solution into the (subtracted) effective action \cite{Huang:2020crf}:
\begin{align}
    S_3' = 4\pi \int_0^\infty dr' r^{\prime 2} \Big[ \Big(\frac{dL}{dr'}\Big)^2 + V'(L)
    -V'(L_+)
    \Big],
    \label{eq:S3prime}
\end{align}
we obtain the values for the classical action $S'_3$.

\subsection{(iv) Interpolating functions}

\begin{figure}[htb]
    \centering
    \includegraphics[width=0.48\linewidth]{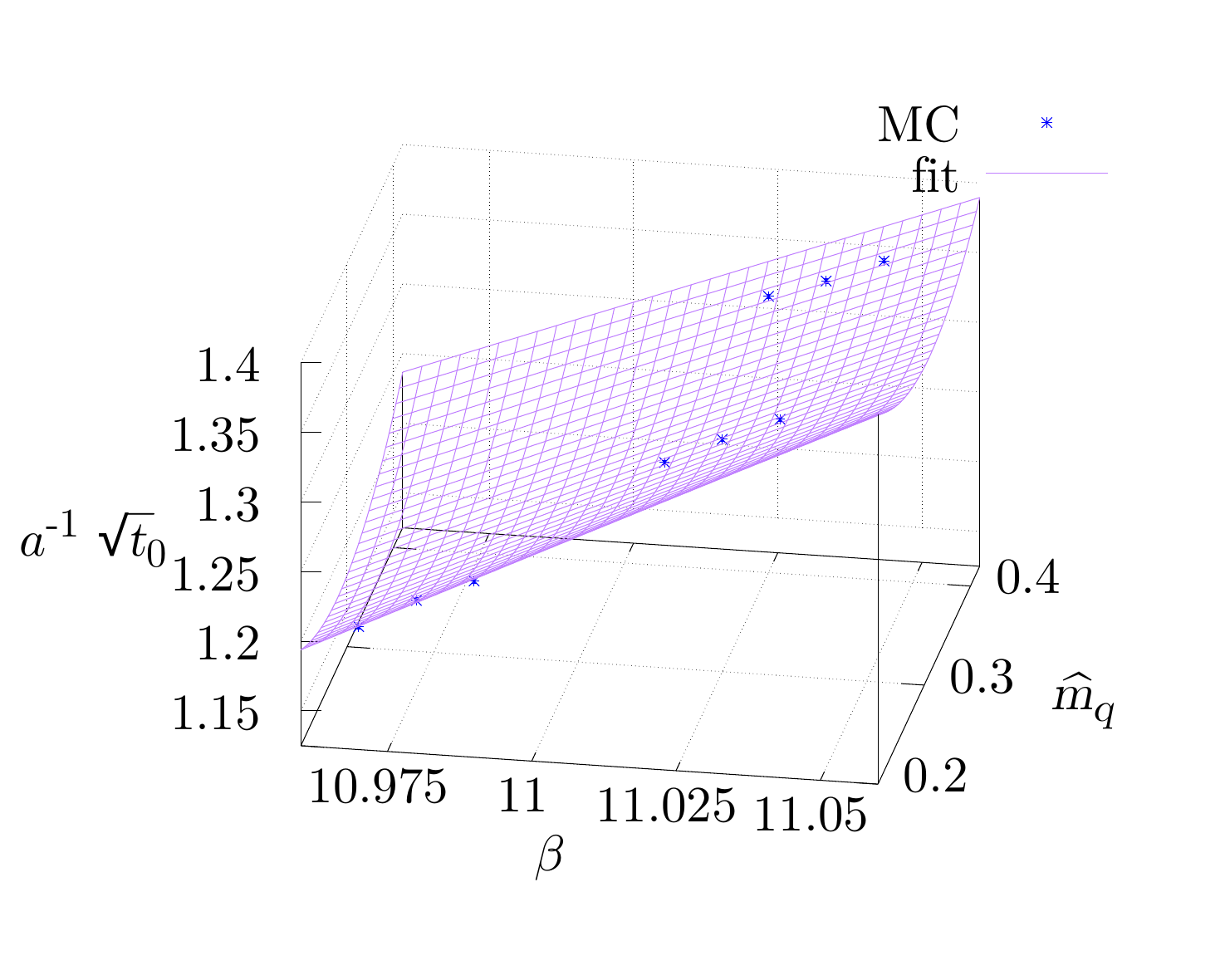}
    \includegraphics[width=0.48\linewidth]{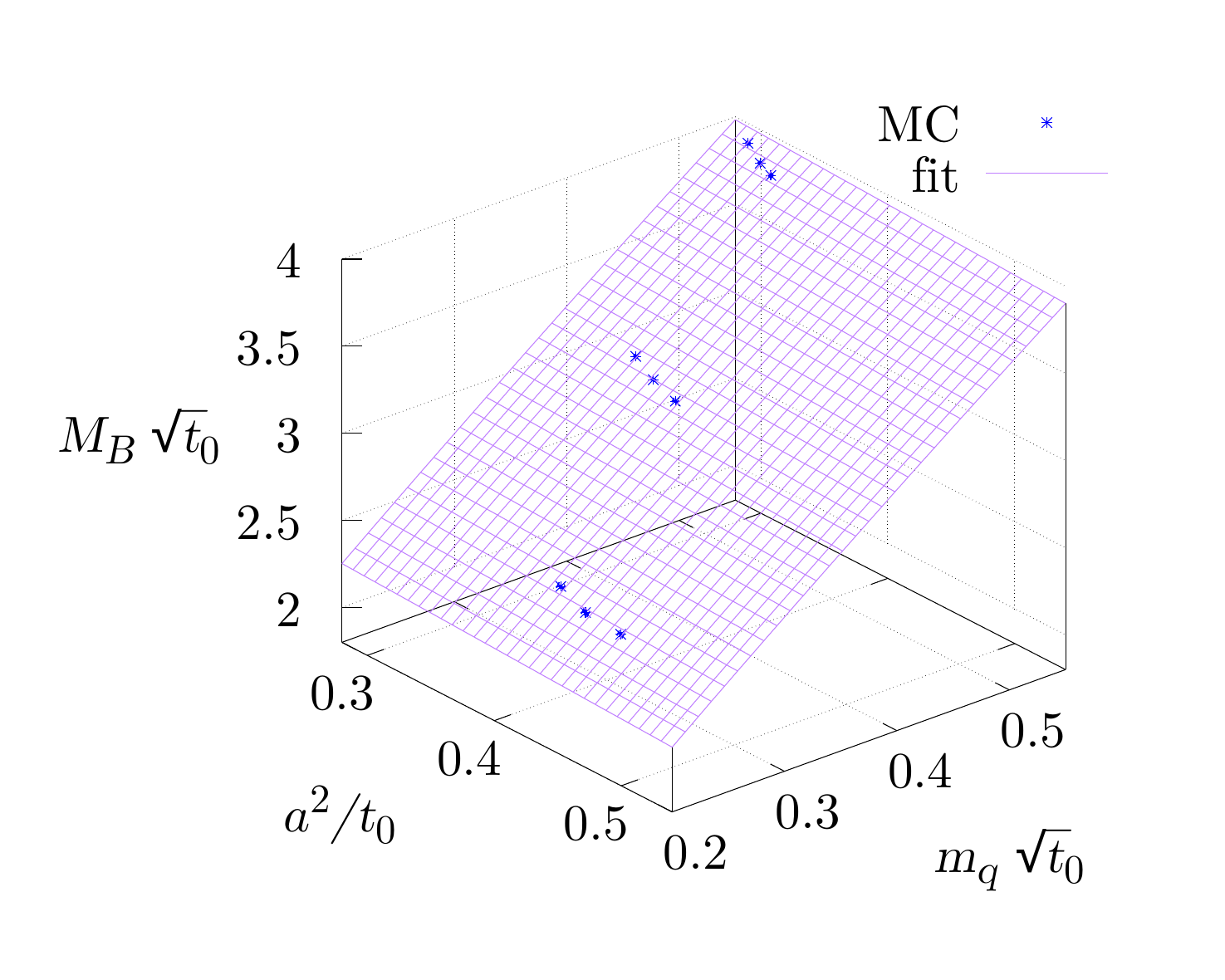}
    \caption{ Global fit for the Monte Carlo (MC) results of (Left) $a^{-1}\sqrt{t_0}$ and (Right) $M_B \sqrt{t_0}$. 
    The reduced $\chi^2$ is 0.94 for $a^{-1} \sqrt{t_0}$ and 2.68 for $M_B \sqrt{t_0}$.}
    \label{fig:globalfit}
\end{figure}

\begin{figure}[htb]
    \centering
    \includegraphics[width=0.53\linewidth]{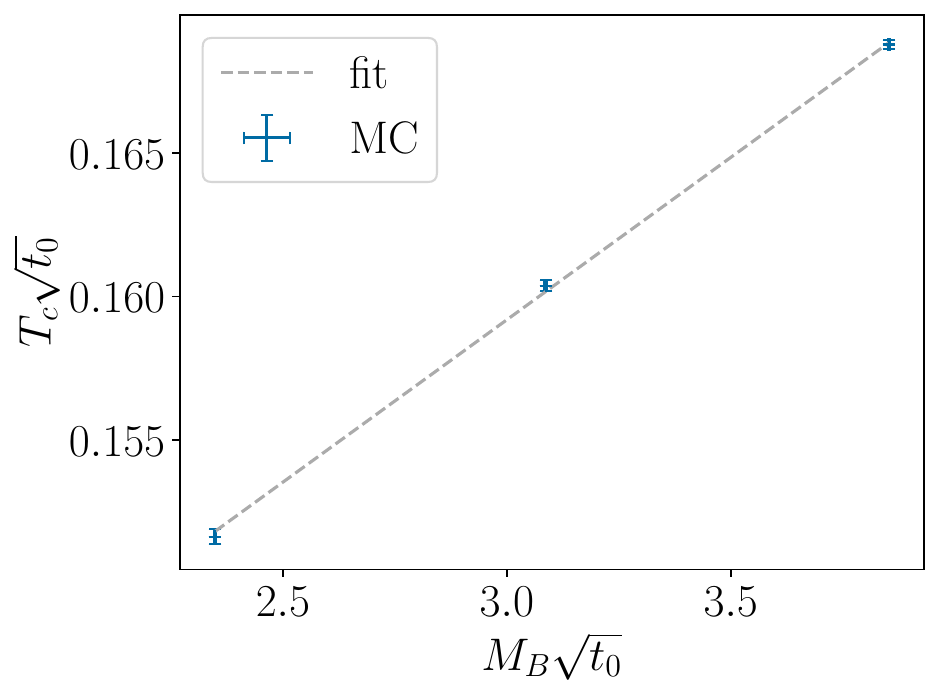}
    \caption{ $T_c \sqrt{t_0}$ fitted with the linear function of $M_B \sqrt{t_0}$. The reduced $\chi^2$ is 2.11. }
    \label{fig:Tc_MB}
\end{figure}

To interpolate the functional dependence of $a^{-1}$ on $\beta$ and $\hat{m}_q$, we use a generic expansion:
\begin{align}
    a^{-1}
    \approx
    c_{00}
    +
    c_{10}\beta
    +
    c_{01} \hat{m}_q
    +
    c_{11} \beta\hat{m}_q
    +
    c_{02} \hat{m}_q^2.
    \label{eq:global_fit_ansatz}
\end{align}
For the baryon mass $M_B$, we use the ansatz:
\begin{align}
    M_B
    \approx
    c_{00} + 
    c_{20} a^2 + 
    c_{01} m_q +
    c_{02} m_q^2
    .
\end{align}
The fit results are shown in Fig.~\ref{fig:globalfit}.
Using the interpolated functions at $\beta_c$ for each $\hat{m}_q$, we obtain the critical temperature $T_c$ at the corresponding $M_B$.
A linear fit is performed to express $T_c$ as a function of $M_B$ (Fig.~\ref{fig:Tc_MB}).

\begin{figure}[htb]
    \centering
    \includegraphics[width=0.48\linewidth]{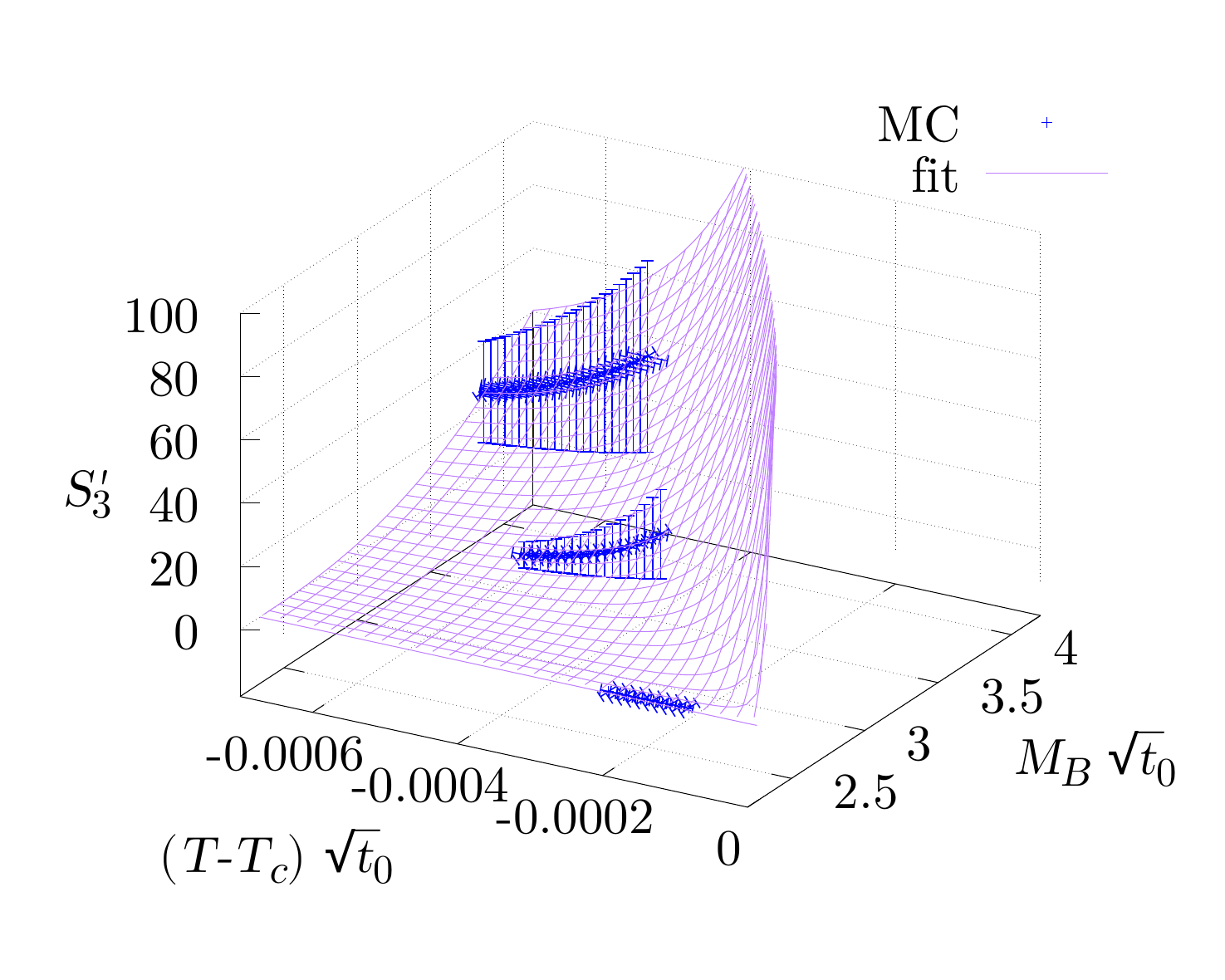}
    \includegraphics[width=0.48\linewidth]{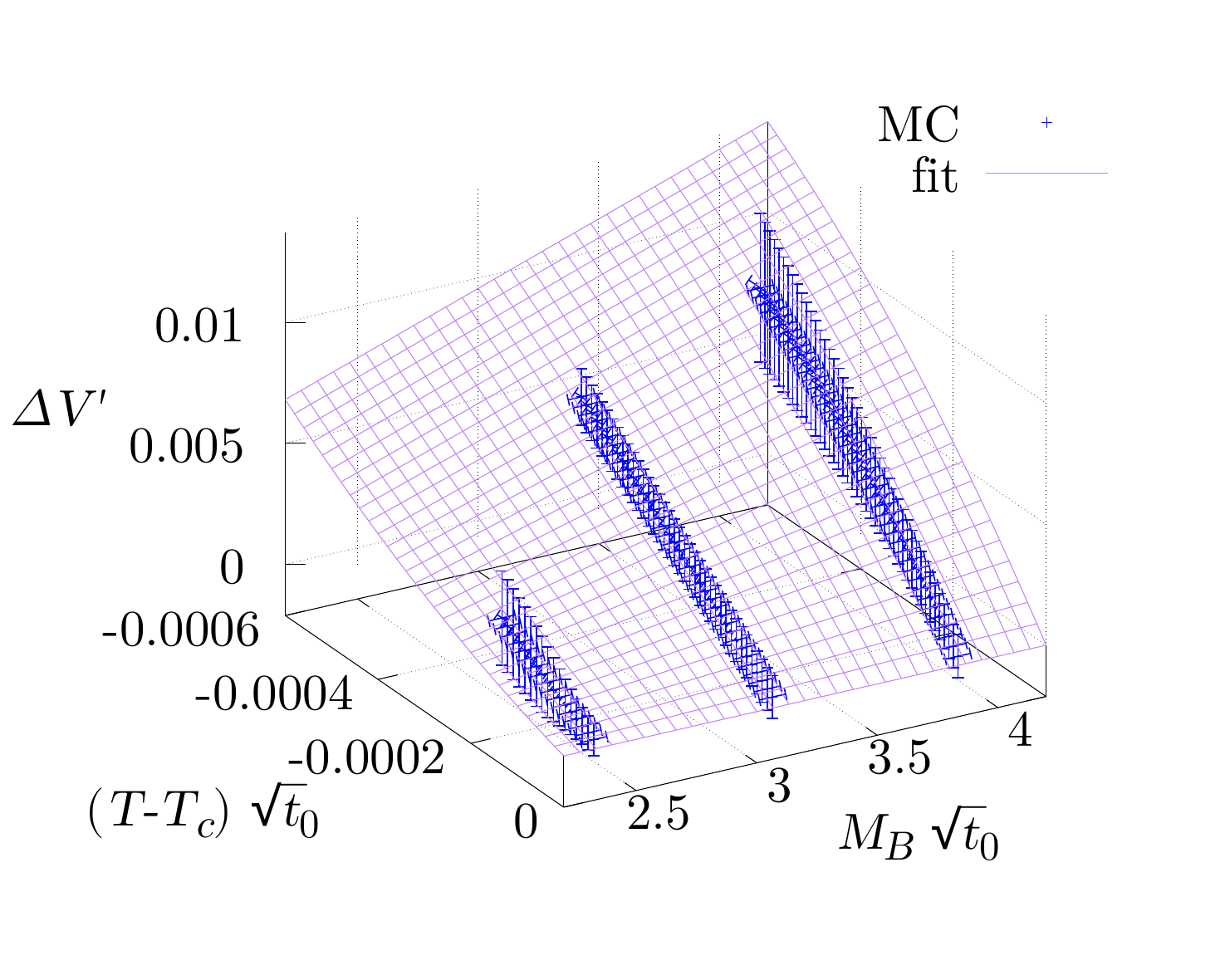}
    \caption{ Interpolating/extrapolating functions for (Left) $S_3' = S_3/T$ and (Right) $\Delta V' = \Delta V / T^4$.
    }
    \label{fig:interpolating}
\end{figure}

Combining the above results, we obtain data points for $S'_3$ and $\Delta V'$ for a given $(T-T_c, M_B)$.
Since the classical action $S'_3$ is quadratically divergent in three dimensions, we use the ansatz with the subleading terms:
\begin{align}
    S'_3 \approx 
    (M_B-M_{B,c})^\gamma 
    \Big( c_0 + \frac{c_1}{T-T_c} + \frac{c_2}{(T-T_c)^2}\Big).
    \label{eq:S3_interpolator}
\end{align}
$\Delta V'$ is interpolated with a generic expansion:
\begin{align}
    \Delta V'
    \approx
    (T-T_c)(b_0 + b_1 M_B) 
    + (T-T_c)^2 ( c_0 + c_1 M_B).
    \label{eq:DeltaV_interpolator}
\end{align}
The fit results are shown in Fig.~\ref{fig:interpolating}.

\subsection{(v) Mass landscape}

\begin{figure}[htb]
    \centering
    \includegraphics[width=0.54\linewidth]{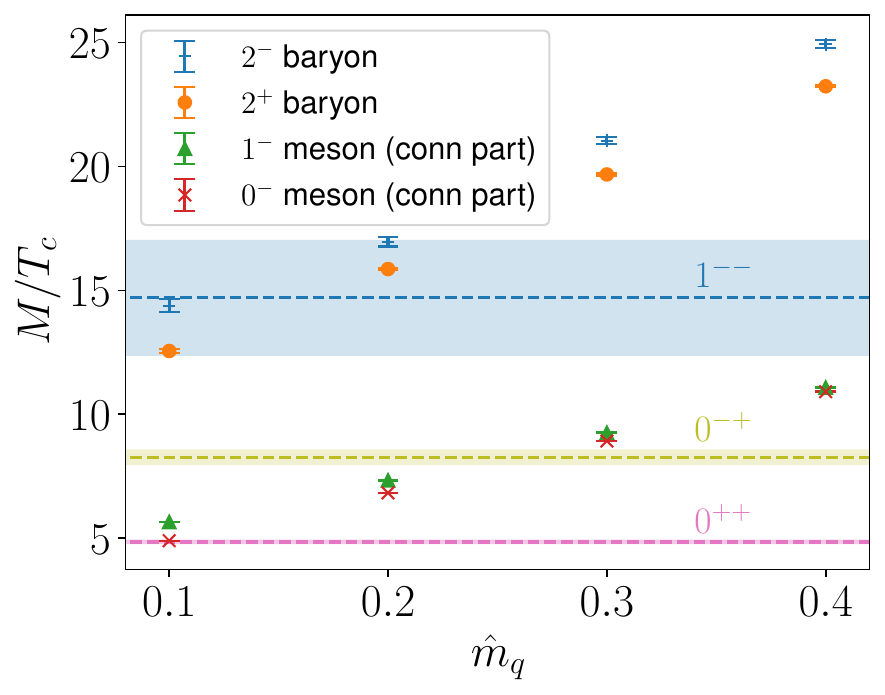}
    \caption{ The lightest masses $M$ in the units of $T_c$ in the connected parts of the baryonic and mesonic correlators, evaluated at $(\hat{m}_q, \beta) = (0.1, 10.865)$, $(0.2, 10.99)$, $(0.3, 11.035)$, and $(0.4, 11.045)$. The lightest quenched glueball masses for a given $J^{PC}$ are indicated by dashed lines, taken from Ref.~\cite{Lucini:2010nv}.
    }
    \label{fig:spec}
\end{figure}

In one-flavor quantum chromodynamics, the confinement scale and the quark mass are the primary scales in the theory.
The mass landscape may be observed through the lightest masses in the connected diagrams for the baryonic and mesonic correlators compared against the quenched glueball masses (Fig.~\ref{fig:spec}).
It should be noted, however, that the mesonic masses may not be regarded as physical due to omitted contributions from the disconnected diagrams.
We nevertheless observe that, roughly around the critical mass $0.1<m_{q,c}<0.2$, the quenched glueball and mesonic masses turn over.
Detailed analysis will be given with full spectroscopy results in future studies.

\onecolumngrid

\section*{Supplementary materials}
\section{Parameter details in lattice calculation}

\begin{figure}[htb]
    \centering
    \includegraphics[width=5.5cm]{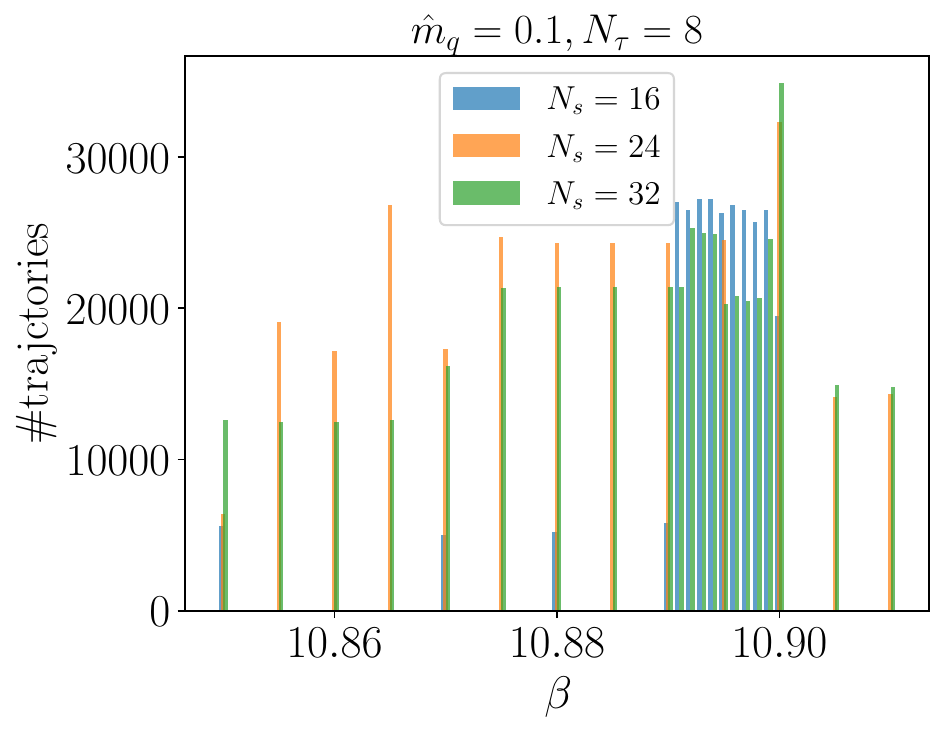}
    \includegraphics[width=5.5cm]{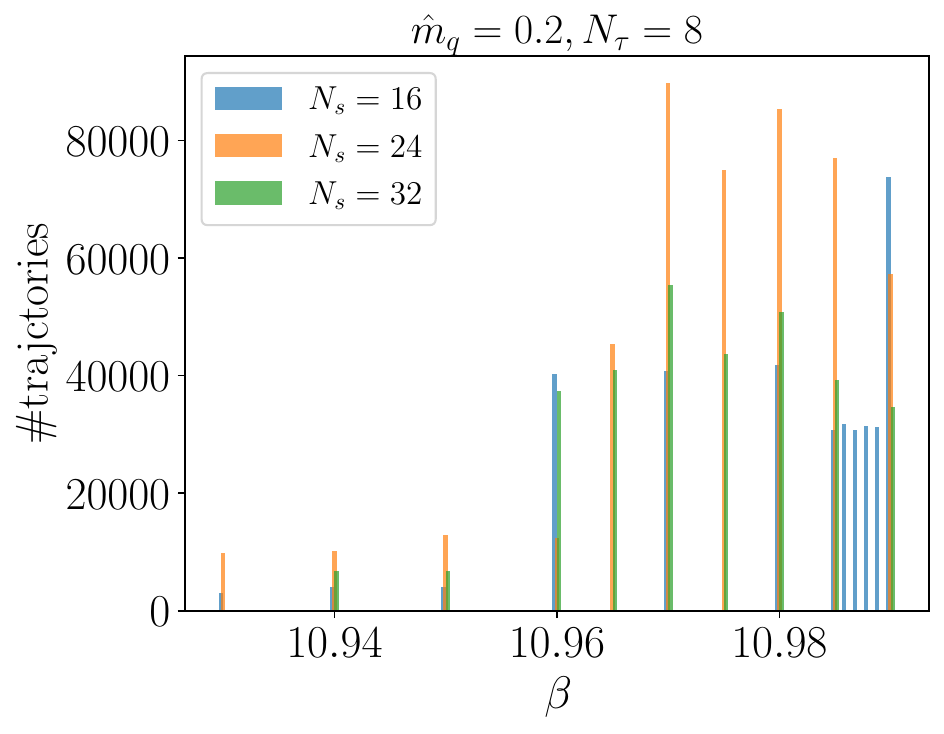}\\
    \includegraphics[width=5.5cm]{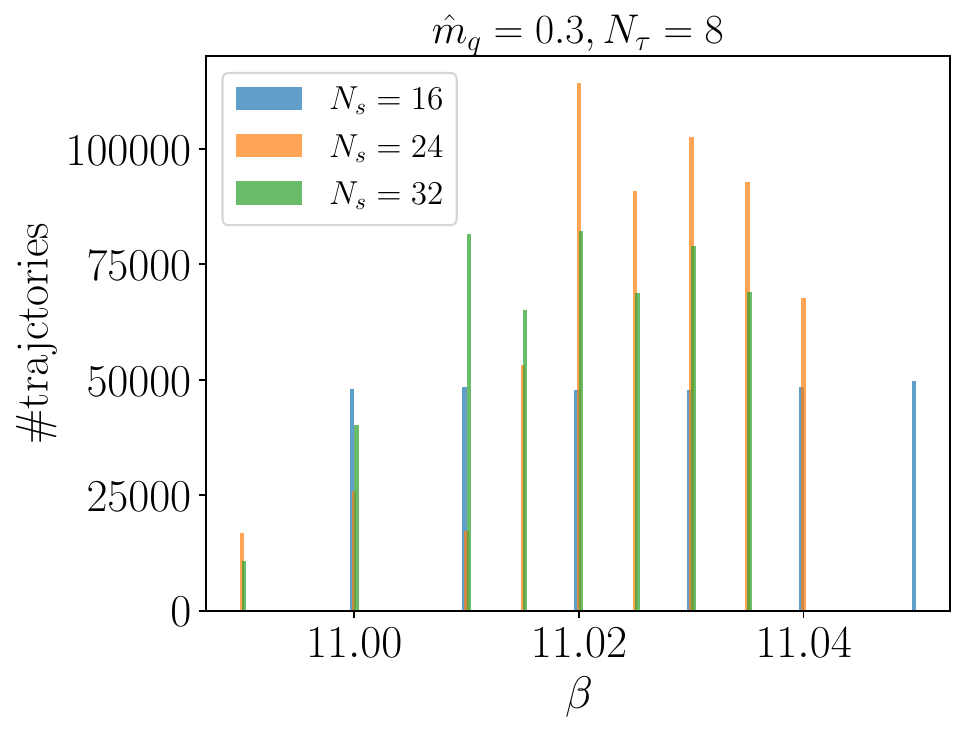}
    \includegraphics[width=5.5cm]{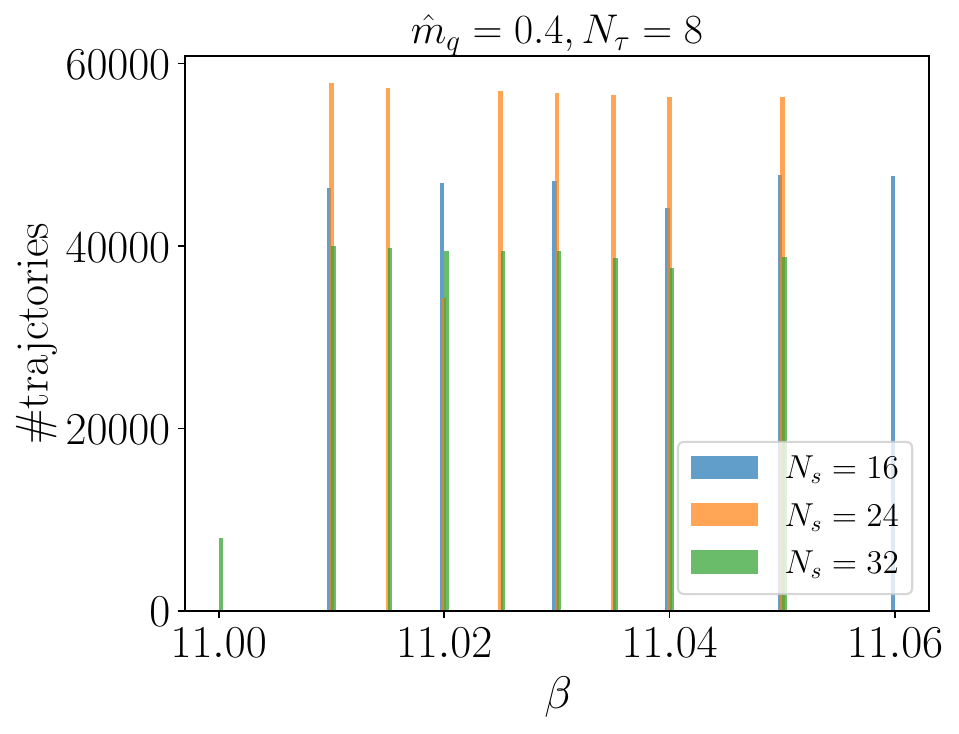}
    \caption{The total number of HMC trajectories after thermalization for the finite temperature ensembles, $\hat{m}_q=0.1,0.2,0.3,0.4$ from top left to bottom right.
    The measurement is performed every 100 trajectories.
    }
    \label{fig:samplesize_FT}
\end{figure}

\begin{figure}[htb]
    \centering
    \includegraphics[width=6cm]{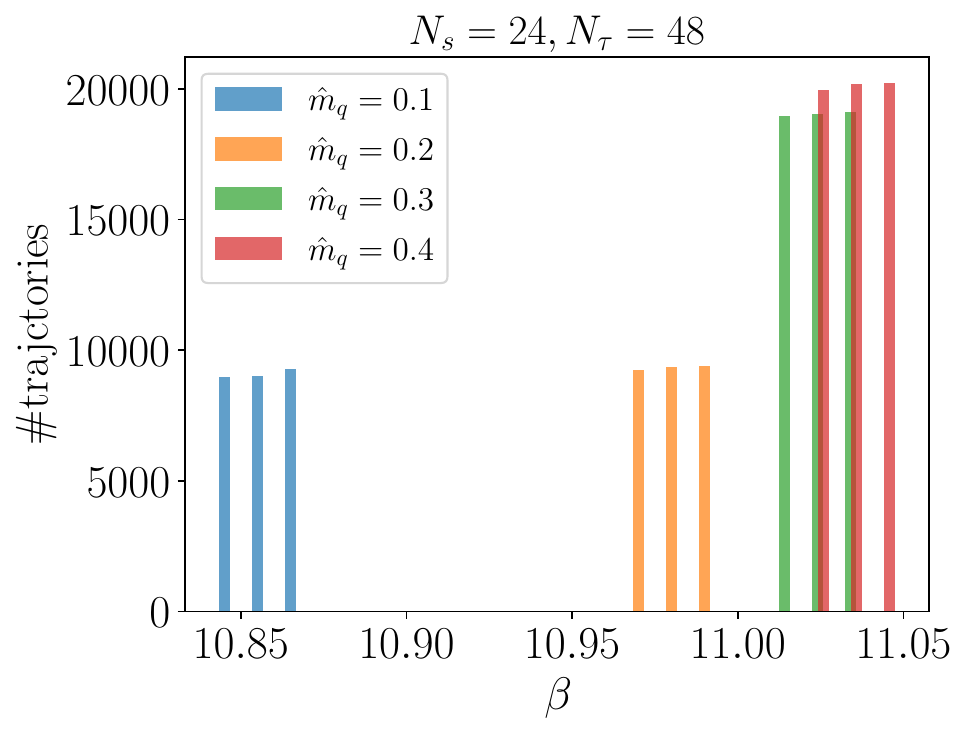}
    \caption{ The total number of HMC trajectories after thermalization for the zero temperature ensembles.
    The measurement is performed every 20 trajectories.
    }
    \label{fig:samplesize_zeroT}
\end{figure}

For the finite-temperature calculation, we create two streams for each parameter $(\hat{m}_q, \beta)$ with the cold and hot starts to ensure thermalization.
The total sample sizes of the generated ensembles are summarized in Fig.~\ref{fig:samplesize_FT}, where the trajectory length in the hybrid Monte Carlo (HMC) algorithm is set to 1.0 in all cases.
To calculate the Wilson flow scale and the baryon mass, we also generate zero-temperature ensembles with $N_s=24$, $N_\tau = 48$.
The total sample sizes of the zero-temperature ensembles are summarized in Fig.~\ref{fig:samplesize_zeroT}.

In the calculation of the histogram, the standard deviations in the $x$ and $y$ directions of the Gaussian regularization of the delta function are chosen to be the same as the widths of the histogram, $\Delta x, \Delta y$.
The histogram parameters are summarized in Table~\ref{tab:histogram_parameter}.
For the multipoint reweighting method, we choose the range $\beta_{\rm min}\leq \beta \leq \beta_{\rm max}$ that includes $\beta_c$ for each $\hat{m}_q$, and the potential is evaluated for the equally spaced $n_\beta$ points in the range.
The reweighting parameters are summarized in Table~\ref{tab:reweighting_parameter}.
Since the multipoint reweighting method combines data from multiple ensembles, the net statistical error is estimated with the error-propagation formula from all ensembles.
The statistical error from each ensemble is calculated by the jackknife method, in which each ensemble is divided into 40 bins.

\begin{table}[htb]
    \centering
    \begin{tabular}{|c|c|c|c|c|c|c|}
    \hline
        $N_s$ & $x_{\rm min}$ & $x_{\rm max}$ & $\Delta x$ & $y_{\rm min}$ & $y_{\rm max}$ & $\Delta y$ \\\hline
        16 & -0.2 & 0.6 & 24 & -0.2 & 0.2 & 24 \\
        24 & -0.2 & 0.6 & 32 & -0.2 & 0.2 & 32 \\
        32 & -0.2 & 0.6 & 40 & -0.2 & 0.2 & 40 \\\hline
    \end{tabular}
    \caption{Histogram parameters}
    \label{tab:histogram_parameter}
\end{table}

\begin{table}[htb]
    \centering
    \begin{tabular}{|c|c|c|c|}
    \hline
         $\hat{m}_q$   & $\beta_{\rm min}$  & $\beta_{\rm max}$  & $n_\beta$ \\\hline
         0.1 & 10.85 & 10.90 & 1000  \\
        0.2 & 10.93 & 10.98 & 1000  \\
        0.3 & 10.995 & 11.045 & 1000  \\
        0.4 & 11.0 & 11.05 & 1000  \\\hline
    \end{tabular}
    \caption{Reweighting parameters. }
    \label{tab:reweighting_parameter}
\end{table}

The critical coupling $\beta_c$ is determined by looking into the discontinuity of the potential minimum $\bar{L}_0$ with the reweighted values of $\beta$ (see Table~\ref{tab:betac}).
Note here that the statistical fluctuation in the lattice data varies both the critical coupling $\beta_c$ ({\it i.e.}, in the horizontal direction) and the value of the potential $V' = V/T^4$ ({\it i.e.}, in the vertical direction).
In performing the jackknife error analysis for $V'$, we separate the two fluctuations by always adjusting $\beta_c$ to be the location of the discontinuity; then the horizontal fluctuation is quoted as the statistical error on $\beta_c$, and the vertical fluctuation is quoted as the statistical error on $V'$.
The potential $V'$ is, in this sense, a function of $\beta-\beta_c$, and thus a function of $(T-T_c)\sqrt{t_0}$ since $a^{-1}\sqrt{t_0}$ is linearized with respect to $\beta$ in our region of interest in Eq.~\eqref{eq:global_fit_ansatz}.

\begin{table}[htb]
    \centering
    \begin{tabular}{|c|c|c|c|c|}
    \hline
        $N_s$\textbackslash $\hat{m}_q$& 0.1 & 0.2 & 0.3 & 0.4 \\\hline
        16 & 10.89630(58) & 10.9444(48) & 10.9985(44) & 11.0103(64) \\
        24 & 10.8859(18) & 10.9717(15) & 11.0172(14) & 11.0366(12) \\
        32 & (10.8681(41)) & 10.9699(11) & 11.02045(89) & 11.04275(76) \\ \hline
    \end{tabular}
    \caption{ Estimates of $\beta_c$.  For $\hat{m}_q=0.1$, $N_s=32$,  $\beta_c$ is determined such that it gives the largest derivative $d\bar{L}_0/d\beta$. }
    \label{tab:betac}
\end{table}

\end{document}